\documentclass[aps,prd,reprint,nofootinbib]{revtex4-1}

\usepackage{amsmath,amssymb,mathtools}
\usepackage{bm}
\usepackage[T1]{fontenc}
\usepackage[utf8]{inputenc}
\usepackage[dvipsnames]{xcolor}
\usepackage{hyperref}
\usepackage{graphicx}
\usepackage{tikz}
\usetikzlibrary{arrows.meta,decorations.markings,decorations.pathmorphing,calc}
\usepackage{microtype}
\usepackage{orcidlink}

\hypersetup{colorlinks=true,linkcolor=blue,citecolor=blue,urlcolor=blue}

\newcommand{\itp}{\affiliation{Institute of Theoretical Physics,
Chinese Academy of Sciences, Beijing 100190, China}}
\newcommand{\ucas}{\affiliation{School of Physical Sciences,
University of Chinese Academy of Sciences, Beijing 100049, China}}
\newcommand{\scnt}{\affiliation{Southern Center for Nuclear-Science Theory (SCNT),
Institute of Modern Physics, Chinese Academy of Sciences, Huizhou 516000, China}}

\begin{document}

\title{Amplitude structure of $3\to 3$ scattering in a Mandelstam variable representation}

\author{Xu Zhang\,\orcidlink{0000-0002-3687-248X}}\email{zhangxu@itp.ac.cn}
\itp

\author{Feng-Kun Guo\,\orcidlink{0000-0002-2919-2064}}
\email{fkguo@itp.ac.cn}
\itp
\ucas
\scnt

\begin{abstract}
We construct a dispersive representation of the relativistic $3\to 3$ scattering amplitude for three identical spinless particles in the $S$-wave. The two-particle subenergy, instead of the total energy, is used as the dispersive variable. This choice keeps the physical dispersive contour free of the kinematical cuts that complicate total-energy dispersion relations. 
By separating discontinuities across the two-particle subenergy cuts from that across the three-body cut, we derive a linear integral equation with one-particle exchange as the driving term. We further show that the solution satisfies three-body unitarity as a consequence of two-body unitarity, analyticity, and crossing symmetry. For pair-wise interactions, this representation can be rewritten into the form used for isobar-spectator scattering. 
Finally, we give prescriptions for contour deformation and the subtraction of poles in the two-body subsystem amplitudes, which continue the amplitude onto adjacent unphysical Riemann sheets and thus provide direct access to the analytic structure relevant for three-body resonance poles.
\end{abstract}

\maketitle

\section{Introduction}
\label{sec:intro}

The hadron spectrum emerges from the internal dynamics of the QCD degrees of freedom. Most hadrons are unstable and are observed in scattering or decay processes with multiparticle final states. Final-state interactions complicate the analysis of the amplitudes of such processes. In addition to usual resonances decaying into three hadrons such as $\omega$~\cite{ParticleDataGroup:2026,Yan:2024gwp} and $a_1(1260)$~\cite{ALEPH:2005qgp,JPAC:2018zwp,Sadasivan:2020syi,Mai:2021nul,Sadasivan:2021emk,Feng:2024wyg}, understanding the nature of a large number of exotic hadrons or exotic candidates requires a careful treatment of three-body dynamics; well-known examples include $a_1(1420)$~\cite{COMPASS:2015kdx,COMPASS:2020yhb,Mikhasenko:2015oxp,Basdevant:2015wma,Aceti:2016yeb}, $\eta(1405/1475)$~\cite{BESIII:2012aa,Nakamura:2022rdd,Nakamura:2023hbt,Cheng:2024sus}, and $\pi_1(1600)$~\cite{E852:1998mbq,COMPASS:2009xrl,COMPASS:2021ogp,JPAC:2018zyd,Dzierba:2005jg} in the light-meson sector, as well as $X(3872)$~\cite{Belle:2003nnu,CDF:2006ocq,Suzuki:2005ha,Baru:2011rs,Schmidt:2018vvl,Braaten:2020nmc,Ji:2025hjw} and $T_{cc}^+$~\cite{LHCb:2021vvq,LHCb:2021auc,Du:2021zzh,Qiu:2023uno,Wang:2023iaz,Dawid:2024dgy,Dawid:2025wsn} in the heavy-hadron sector. To properly extract the properties of these hadrons from experimental data, it is necessary to construct amplitudes for three-particle scattering that satisfy the fundamental principles of unitarity and analyticity in quantum field theory.

Faddeev equations provide a nonrelativistic setting for three-body scattering~\cite{Faddeev:1960su,faddeev:1965ml}. The scattering amplitude can be obtained as the solution of a system of coupled linear integral equations once the potential describing the particle interactions is given. However, a nonrelativistic treatment is not adequate in the domain of high-energy particle reactions. In the relativistic regime, the construction of the $3\to3$ scattering amplitude from the $S$-matrix constraints of unitarity and analyticity has been investigated extensively over the past half century~\cite{Blankenbecler:1961zz,Cook:1962zz,Fleming:1964zz,Mandelstam:1965xx,Grisaru:1966uev,Blankenbecler:1965gx,Aaron:1968aoz,Aaron:1973ca,Amado:1974za,osti_5335708} and also in recent investigations~\cite{Mai:2017vot,Mai:2017bge,Jackura:2018xnx,Mikhasenko:2019vhk,Jackura:2019bmu,Dawid:2020uhn}. 

In Ref.~\cite{Mandelstam:1965xx}, the $3\to3$ scattering amplitude with a generalized two-particle subsystem interaction, not factorized into pairwise two-body amplitudes, is formulated through an $N/D$ equation. Besides the total angular momentum and the helicities of the initial and final states, there are five independent energy variables: the total energy $s$, two initial-state subenergies, and two final-state subenergies. 
For the fully relativistic problem, Mandelstam replaces the subenergies by the dimensionless variables $x_i=\bm{p}_i^2/(\bm{p}_1^2+\bm{p}_2^2+\bm{p}_3^2)$, and analogously $x_i'$ for the final state, where $\bm{p}_i$ is the three-momentum of the $i$th incoming particle in the overall center-of-mass (c.m.) frame. 
At fixed $x_i$, the subenergies are multivalued functions of $s$. The squared total energy is used as the dispersion variable, which, however, introduces kinematical branch points. These kinematical singularities persist even for spinless particles after partial-wave decomposition and complicate the analytic structure. 
The function entering the $N/D$ dispersion equation lies on a Riemann surface divided into multiple sheets by these singularities. The dynamical discontinuity can be determined on each kinematical sheet following Refs.~\cite{Olive:1965xy,Eden:1966dnq}, whereas the discontinuity across a kinematical cut is not known. This makes three-particle scattering substantially more complicated than two-particle scattering~\cite{Mandelstam:1965xx}. 
Related kinematical singularities in two-particle scattering with spin have also been studied extensively~\cite{Cohen-Tannoudji:1968lnm,Jackson:1968rfn,Franklin:1966qrp,Wang:1966zza,Hara:1964zza,Stoica:2011cy,Lutz:2011xc}.

In Refs.~\cite{Fleming:1964zz,Grisaru:1966uev}, the explicit form of the discontinuity relations for $3\to 3$ scattering is derived for all relevant energy variables. 
By parameterizing the two-particle subsystem via pairwise interactions, the two-particle scattering amplitude can be factored out of the three-particle scattering amplitude. The $N/D$ formalism is constructed from the amputated amplitude. 
This formalism is simpler than that given in Ref.~\cite{Mandelstam:1965xx}, where the two-particle subsystem interaction is in a general form without the pairwise parameterization. The three-particle discontinuity relations were reexamined in Refs.~\cite{Mikhasenko:2019vhk,Jackura:2018xnx}. 
Alongside the $B$-matrix parameterization, an effective Bethe--Salpeter construction was given in Ref.~\cite{Aaron:1968aoz} and developed further in Ref.~\cite{Mai:2017vot}, which derived a relativistic isobar--spectator equation that satisfies two- and three-body unitarity. In the physical region, these constructions have the same unitarity discontinuity, whereas their treatment of the unphysical subenergy region and hence their integration bounds can differ~\cite{Mai:2017vot,Dawid:2020uhn,Jackura:2019bmu}. 
The isobar--spectator construction is extended to finite volume in Ref.~\cite{Mai:2017bge}, and its relation to other infinite-volume three-particle scattering formalisms is analyzed in Ref.~\cite{Jackura:2019bmu}.

The purpose of this work is to construct a Mandelstam-variable representation of $3\to 3$ scattering. 
Following Refs.~\cite{Fleming:1964zz,Grisaru:1966uev}, the unitarity relation for three-particle scattering is decomposed into individual contributions from crossing the cuts in the two-particle subenergies and crossing the cut in the total energy. 
The $3\to 3$ scattering amplitude is obtained from dispersion relations using the subenergy as the dispersive integral variable. The construction is motivated by the Khuri-Treiman equation for the $1\to 3$ decays~\cite{Khuri:1960zz}, where the amplitude is based on two-particle unitarity relations. As discussed in Refs.~\cite{Aitchison:1966lpz,Pasquier:1968zz}, the inhomogeneous term guarantees that the amplitude satisfies three-body unitarity. Three-body unitarity is automatically generated by two-body unitarity, analyticity, and crossing symmetry.

The $3\to3$ scattering amplitude appears implicitly in the context of the $1\to3$ process. The relationship between the Faddeev and Khuri--Treiman equations for these two processes has been discussed in Refs.~\cite{Aitchison:1965zz,Aitchison:1966lpz,Pasquier:1968zz}. 
The Khuri--Treiman terms can be related to the corresponding Feynman terms summed in the Faddeev equation with a zero-range potential. In both approaches, the lowest-order nontrivial term corresponds to one real-particle exchange (RPE). 
The analytic properties of the $1\to3$ process in the dispersive approach have been investigated in Refs.~\cite{Barton:1961xw,Aitchison:1965xx,Kacser:1966xx,Bronzan:1963mby}. 
To derive the decay amplitude on the physical Riemann sheet, the precise continuation path is emphasized in Refs.~\cite{Bronzan:1963mby,Hwa:1964ujn,Kacser:1963zz}. We follow closely that approach and analyze the integration contours that place the $3\to3$ amplitude on the physical and adjacent unphysical Riemann sheets.

The present work is organised as follows.
Section~\ref{sec:notation} collects the kinematic notation.
The unitarity condition is discussed in Section~\ref{sec:analytic}.
Section~\ref{sec:disrep-full} sets up the dispersive representation of the $3\to3$ scattering amplitude, and Section~\ref{sec:disrep-pair} introduces the pairwise form. 
Section~\ref{sec:demonst} relates the parameterization in terms of pairwise interactions to other parameterizations.
In Section~\ref{sec:continua}, we discuss the analytic continuation of the $3\to 3 $ scattering amplitude.
Section~\ref{sec:summary} summarises the results.

\section{Kinematics and amplitude}
\label{sec:notation}
We consider the case of three identical spinless particles in an isoscalar state. A one-particle state is normalized as
\begin{equation}
\langle \vec{p}\,'|\vec{p}\,\rangle =2E(2\pi)^3\delta^3( \vec{p}\,'-\vec{p}\,)\equiv\widetilde{\delta}( \vec{p}\,'-\vec{p}\,).
\end{equation}
The symmetrized two-particle state is defined as
\begin{equation}
|\vec{p}_1\,\vec{p}_2\,\rangle =\frac{1}{2}\left(  |\vec{p}_1\,\rangle| \vec{p}_2\,\rangle + |\vec{p}_2\,\rangle| \vec{p}_1\,\rangle\right),
\label{eq:2particlestate}
\end{equation}
and the three-particle state is
\begin{align}
|\vec{p}_1\,\vec{p}_2\,\vec{p}_3\,\rangle &=\frac{1}{3!}\left(  |\vec{p}_1\,\rangle |\vec{p}_2\,\rangle |\vec{p}_3\,\rangle + \text{symm.}\right)\nonumber \\
&\equiv \frac{1}{3}\left(  |\vec{p}_1\,(\vec{p}_2\,\vec{p}_3)\,\rangle + |\vec{p}_2\,(\vec{p}_3\,\vec{p}_1)\,\rangle + |\vec{p}_3\,(\vec{p}_1\,\vec{p}_2)\,\rangle\right), \label{eq:3particlestate}
\end{align}
where symm. denotes all possible symmetric permutations, and the two-particle states in parentheses are the symmetrized states.
\begin{figure*}
   \includegraphics[width=0.8\textwidth]{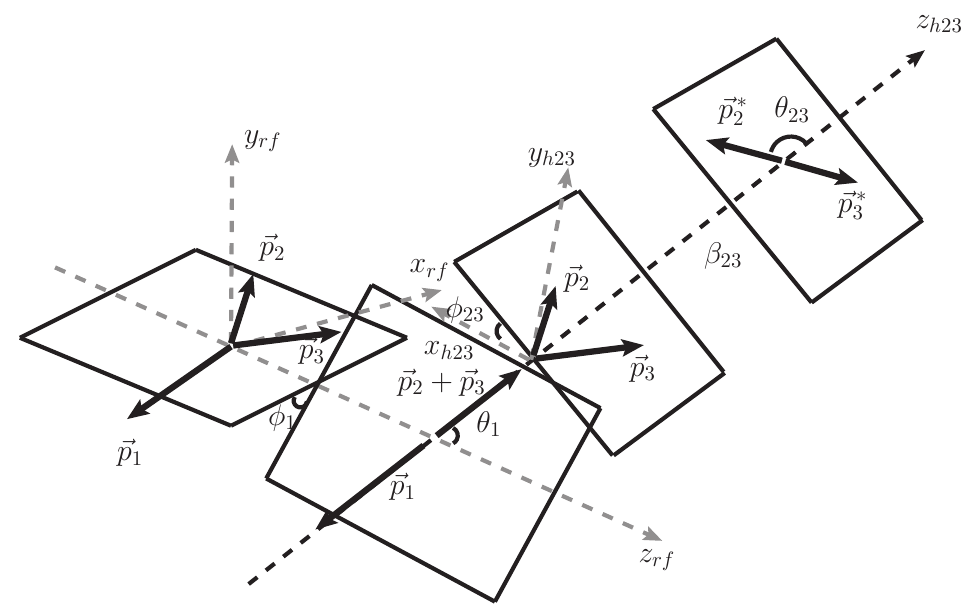}
  \caption{Definition of angles that parametrize three-body kinematics with respect to the given reference frame (see also Ref.~\cite{Mikhasenko:2019vhk}). }
  \label{fig:tbframe}
\end{figure*}

The unit operator in the symmetrized three-particle space can be expressed as
\begin{align}
I_3=&\int d\vec{p} \,|\vec{p}_1\,\vec{p}_2\,\vec{p}_3\,\rangle\langle\vec{p}_1\,\vec{p}_2\,\vec{p}_3|\,=\frac{1}{3}\int d\vec{p} |\vec{p}_1\,(\vec{p}_2\,\vec{p}_3)\,\rangle \nonumber \\
&\times\langle \vec{p}_1\,(\vec{p}_2\,\vec{p}_3)\,|+\frac{2}{3}\int d\vec{p} |\vec{p}_1\,(\vec{p}_2\,\vec{p}_3)\,\rangle\langle \vec{p}_2\,(\vec{p}_3\,\vec{p}_1)\,|\,,
\end{align}
where the integral measure is 
\begin{align}
d\vec{p}\equiv \frac{d^3p_1}{2E_1(2\pi)^3}\frac{d^3p_2}{2E_2(2\pi)^3}\frac{d^3p_3}{2E_3(2\pi)^3}\,.
\end{align}
The projector onto the fully symmetrized three-particle space contains
nine terms. Since the measure $d\vec p$ is invariant under relabeling
of the dummy integration variables, the three diagonal terms are equal, and the six off-diagonal terms are also equal.

The symmetrized amplitude is
\begin{align}
\langle \vec{p}\,'\!\!_{1'}\vec{p}\,'\!\!_{2'}\vec{p}\,'\!\!_{3'}|&T|\vec{p}_1\,\vec{p}_2\,\vec{p}_3\,\rangle\nonumber \\
=&\,\frac{1}{9} 
\sum_{i,j}\langle (\vec{p}\,'\!\!_{j1}\vec{p}\,'\!\!_{j2})\vec{p}\,'\!\!_{j3}|T|(\vec{p}_{i1}\vec{p}_{i2})\,\vec{p}_{i3}\,\rangle, 
\end{align}
where the two-particle states in parentheses are the symmetrized states in Eq.~\eqref{eq:2particlestate}.

The partial-wave decomposition of $\langle (\vec{p}\,'\!\!_{2'}\vec{p}\,'\!\!_{3'})\vec{p}\,'\!\!_{1'}|T|(\vec{p}_2\vec{p}_3)\,\vec{p}_1\,\rangle$ is defined by
\begin{align}
\langle (\vec{p}&\,'\!\!_{2'}\vec{p}\,'\!\!_{3'})\vec{p}\,'\!\!_{1'}|T|(\vec{p}_2\vec{p}_3)\,\vec{p}_1\,\rangle=\sum_{l'\lambda'}\sum_{l\lambda}(2l'+1)^{1/2}D_{\lambda'0}^{l*}(\Omega_{23}')\nonumber\\
&\times(2l+1)^{1/2}D_{\lambda0}^l(\Omega_{23})\langle\vec{p}\,'\!\!_{1'}|\langle \vec{q}\,'\!\!_{1'}l'\lambda'|T|\vec{q}_1l\lambda\,\rangle|\vec{p}_1\,\rangle\,,
\end{align}
where the two-particle state $|\vec{p}_2\vec{p}_3\rangle$ defined in Eq.~\eqref{eq:2particlestate} has the following partial-wave expansion using the Wigner $D$ function,
\begin{align}
|\vec{p}_2\vec{p}_3\rangle=\sum_{l\lambda}(2l+1)^{1/2}D_{\lambda0}^l(\Omega_{23})|\vec{q}_1l\lambda\,\rangle\,.
\end{align}
The spherical angles $\Omega_{23} = (\theta_{23}, \phi_{23})$ specify the direction of the momentum $\vec p_2$ in the rest frame (denoted as ${h23}$) of the pair $(23)$, as shown in Fig.~\ref{fig:tbframe} (see also Ref.~\cite{Mikhasenko:2019vhk}).
We stress that the direction of the $z_{h23}$ axis is chosen such that the spin projection $\lambda$ is conserved and becomes helicity when the state $|\vec{q}_1l\lambda\,\rangle$ is boosted from the $h_{23}$ frame to the production frame. The total 3-momentum of this pair is denoted as $\vec{q}_1=\vec{p}_2+\vec{p}_3$.

Analogously, the partial-wave state $|P,jml\lambda\,\rangle_1$ of the two-particle state and the spectator is obtained by 
projecting the state $|\vec{q}_1\,\rangle|\vec{p}_1\,\rangle$ by the Wigner-$D$ function,
\begin{align}
|\vec{q}_1l\lambda\,\rangle|\vec{p}_1\,\rangle=\sum_{jm}(2j+1)^{1/2}D_{m\lambda}^j(\Omega_{1})|P,jml\lambda\,\rangle_1\,,
\end{align}
where $\vec{P} = \vec{p}_1+\vec{p}_2+\vec{p}_3$ is the total momentum of the three-particle system, $j$ and $m$ are the total orbital momentum and its projection to the $z_{rf}$ axis, respectively,
and $\Omega_1$ represents the spherical angles of the vector $\vec q_1$ in the reference frame as shown in Fig.~\ref{fig:tbframe}.

The partial-wave decomposed amplitude can be written as 
 \begin{align}
 & \langle (\vec{p}\,'\!\!_{2'}\vec{p}\,'\!\!_{3'})\vec{p}\,'\!\!_{1'}|T|(\vec{p}_2\vec{p}_3)\,\vec{p}_1\,\rangle= \sum_{j'm'}\sum_{jm}\sum_{l'\lambda'}\sum_{l\lambda} 
 \sqrt{2j'+1} \nonumber \\
&\times D^{j'*}_{m'\lambda'}(\Omega'_1)\sqrt{2j+1}D^{j}_{m\lambda}(\Omega_1)
  \sqrt{2l'+1} D^{l'*}_{\lambda'0}(\Omega'_{23})
  \nonumber \\
&\times \sqrt{2l+1}D^{l}_{\lambda0}(\Omega_{23}){}_{1'}\langle P',j'm'l'\lambda'|T|P,jml\lambda\rangle_1\nonumber \\
&\times (2\pi)^4\delta^{(4)}(P'-P)\,
  \delta_{j'j}\delta_{m'm}\,.
\end{align}

We will refer to $\sigma$ defined as
\begin{equation}
   \sigma \equiv q^2= (p_2+p_3)^2,
\end{equation}
as the corresponding two-body subenergy variable, and its
positive square root is the pair c.m. energy.
Analogously, the final-state subenergy variable is
\begin{equation}
   \sigma' \equiv q'^2= (p_2'+p_3')^2,
\end{equation}
while the total four-momentum squared is
\begin{equation}
   s = P^2.
\end{equation}

A disconnected contribution arises when one particle propagates as a spectator
from the initial to the final state while the remaining two particles undergo
two-body scattering. Accordingly, the full $3\to3$ amplitude is decomposed as
$T=T_d+T_c$, where $T_d$ and $T_c$ denote the disconnected and connected parts,
respectively.
The disconnected amplitude is 
\begin{align}
\langle \vec{p}\,'\!\!_1\vec{p}\,'\!\!_2\vec{p}\,'\!\!_3|T_d|\vec{p}_1\,\vec{p}_2\,\vec{p}_3\,\rangle =&\,\frac{1}{3}\sum_{i,j} \Tilde{{\delta}}(\vec{p}\,'\!\!_{j3}-\vec{p}_{i3})\nonumber\\
&\times \langle \vec{p}\,'\!\!_{j1}\vec{p}\,'\!\!_{j2}|t|\vec{p}_{i1}\vec{p}_{i2}\,\rangle,
\end{align}
where $t$ is the two-particle scattering amplitude, the factor $3$ indicates three possible ways to choose the noninteracting spectator particle,
and the connected part is
\begin{align}
\langle \vec{p}\,'\!\!_{1'}\vec{p}\,'\!\!_{2'}\vec{p}\,'\!\!_{3'}|&T_c|\vec{p}_1\,\vec{p}_2\,\vec{p}_3\,\rangle  \nonumber \\
=&\,\frac{1}{9} 
\sum_{i,j}\langle (\vec{p}\,'\!\!_{j1}\vec{p}\,'\!\!_{j2})\vec{p}\,'\!\!_{j3}|T_c|(\vec{p}_{i1}\vec{p}_{i2})\,\vec{p}_{i3}\,\rangle.
\end{align}

\section{Analyticity and unitarity}
\label{sec:analytic}

In what follows, we consider three identical spinless particles of common mass $m_\pi$, and isospin degrees of freedom are neglected.
The unitarity condition for the $3\to3$ scattering amplitude takes a particularly simple form when the total squared energy $s$ and the pair subenergy variables $\sigma^{(')}$ are used as kinematic variables. 
In addition to the ordinary right-hand cuts, the amplitude has discontinuities associated with crossed channels. 
We first consider the unitarity relation for two-body-subenergy discontinuity.
The three-body phase space is
\begin{align} 
    d \Phi_3 &= \frac{d^3 p_1''}{2E_1''(2\pi)^3} \frac{d^3 p_2''}{2E_2''(2\pi)^3} \frac{d^3 p_3''}{2E_3''(2\pi)^3} (2\pi)^4\delta^4(P''-P) \nonumber \\
    & = \frac{d \sigma_1''}{2\pi}\,\rho(\sigma_1'') \rho_s(\sigma_1'') \frac{d \Omega_1}{4\pi}\frac{d \Omega_{23}}{4\pi}\,
    \theta^+(\phi(\sigma_2'',s,\sigma_3'')) \nonumber\\
    & = \frac{d \sigma_2'' d \sigma_3''}{2\pi(8\pi)^2 s} \frac{d \Omega_2}{4\pi}\frac{d \phi_{31}}{2\pi}\,
    \theta^+(\phi(\sigma_2'',s,\sigma_3'')),
\end{align}
where $\rho(\sigma) = \lambda^{1/2}(\sigma,m_\pi^2,m_\pi^2)/(8\pi\sigma)$, and $\rho_s(\sigma)={\lambda_s^{1/2}(\sigma)}/{(8\pi s)}$, with $\lambda(x,y,z) = x^2+y^2+z^2-2xy-2yz-2zx$ being the K\"all\'en function and $\lambda_s(\sigma)\equiv\lambda(s,\sigma,m_{\pi}^2)$.
The function $\theta^+(\phi(\sigma_2,s,\sigma_3))$ restricts the variables to the physical domain of the three-body phase space:
\begin{align}\label{eq:Heaviside.plus}
  \theta^+(\phi(\sigma_2,s,\sigma_3)) \equiv &\,\theta(\phi(\sigma_2,s,\sigma_3)) \,\theta(s-9m_{\pi}^2) \nonumber \\
  &\times\theta(\sigma_2-4m_{\pi}^2)\theta(\sigma_3-4m_{\pi}^2)\,,
\end{align}
with $\phi$ being the Kibble function, 
\begin{align}
  \phi(\sigma_2,s,\sigma_3) = \sigma_2\sigma_3(3m_{\pi}^2+s-\sigma_2-\sigma_3)-m_{\pi}^2(s-m_{\pi}^2)^2,
\end{align} 
where the factors
$\theta(s-9m_\pi^2)$ and $\theta(\sigma_i-4m_\pi^2)$ $(i=2,3)$ impose the three- and
two-particle thresholds, respectively, while $\theta(\phi)$ enforces $\phi\geq 0$.
\begin{figure}
   \includegraphics[width=0.45\textwidth]{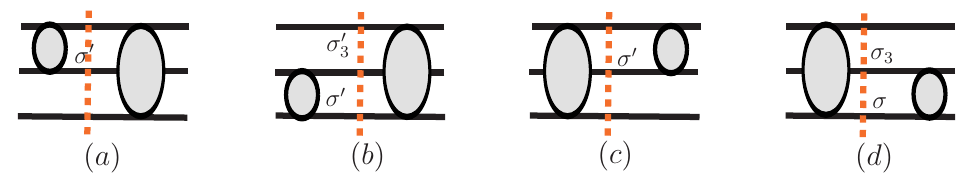}
   \includegraphics[width=0.45\textwidth]{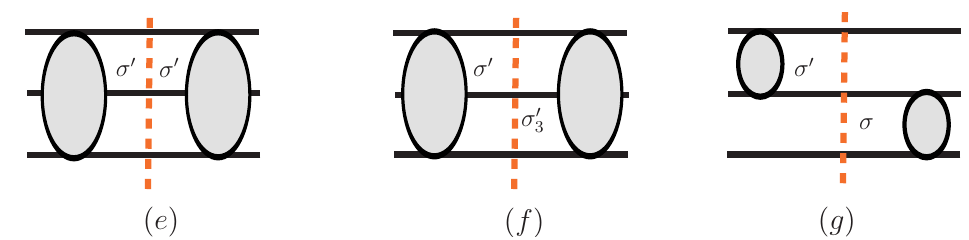}
  \caption{ Diagrammatic representation for the $3\to3$ connected amplitude unitarity relation.}
  \label{fig:dis-gro}
\end{figure}

We now consider the unitarity relation generated by intermediate states with disconnected-connected ($dc$), connected-disconnected ($cd$) and disconnected-disconnected ($dd$) topologies.
The contribution from the two-particle subenergy cuts can be written as 
~\cite{Fleming:1964zz,Olive:1965xy,Eden:1966dnq,Hwa:1964ujn}
\begin{align} 
\label{eq:unit-ral}
  &[\langle{b}|T_c-T_c^\dagger|{a}\rangle]_{dc+cd+dd} = \nonumber\\
  &i\int d \Phi_3 \big[
\langle{b}|T_d^\dagger|{1''}\rangle\langle{1''}| T_c|{a}\rangle + 2\langle{b}|T_d^\dagger|{2''}\rangle\langle{3''}|T_c|{a}\rangle     \nonumber \\
     &+ \langle{b}|T_c^\dagger|{1''}\rangle\langle{1''}| T_d|{a}\rangle + 2\langle{b}|T_c^\dagger|{2''}\rangle\langle{3''}|T_d|{a} \rangle    \nonumber\\
     &+ 6\langle{b}|T_d^\dagger|{2''}\rangle\langle{3''}|T_d|{a}\rangle\big],
\end{align}
where we introduced a compact notation for the state $|{a} \rangle= |{p_{a_1}}\rangle|{p_{a_2}p_{a_3}}\rangle$,
symmetrized over the momenta $p_{a_2}$, $p_{a_3}$,
e.g., $|{1}\rangle =|{p_{1}}\rangle|{p_{2}p_{3}}\rangle$. For simplicity, we will only consider the $S$-wave projection, $j=l=0$.

The disconnected amplitude is 
\begin{align} 
&\langle{1'}| T_d|1\rangle =(2\pi)^4\delta^4(q\,'_1-q_1)\,\Tilde{{\delta}}( \vec{p}\,'_1-\vec{p}_1\,)\,t(\sigma_1).
\end{align}
Using the relation 
\begin{align}
 \Tilde{{\delta}}( \vec{p}\,'_1-\vec{p}_1\,)  &=(2\pi)^3 2E_1 \delta^3(\vec p\,'_1-\vec p_1) \nonumber \\
 &   = \frac{2\pi}{\rho_s(\sigma_1)}
    \delta(\sigma_1'-\sigma_1)\,4\pi\delta(\Omega_1'-\Omega_1),
\end{align}
the partial-wave amplitude is 
\begin{align} 
&\langle P',0000| T_d|P,0000\rangle \nonumber \\
&=(2\pi)^4\delta^4(P'-P)\,(2\pi)\delta(\sigma'-\sigma)\frac{t(\sigma)}{\rho_s(\sigma)}.
\end{align}
The partial-wave projection of the connected amplitude is
\begin{align} 
&\langle P',0000| T_c|P,0000\rangle =(2\pi)^4\delta^4(P'-P)\,T(\sigma',s,\sigma).
\end{align}

The unitarity relation can be obtained by replacing $|{a} \rangle$ and $|{b} \rangle$ by the projected states $|P,0000 \rangle_a$ and $|P,0000 \rangle_b$. Using
\begin{align} 
&\langle{1''}| T_c|P,0000\rangle =(2\pi)^4\delta^4(P''-P)\,T(\sigma_1'',s,\sigma),\nonumber \\
&\langle{1''}| T_d|P,0000\rangle =(2\pi)^4\delta^4(P''-P)\,(2\pi)\delta(\sigma_1''-\sigma)\frac{t(\sigma)}{\rho_s(\sigma)},
\end{align}
and inserting 
$\langle \vec{p}\,'_{1'}|\vec{p_1}\,\rangle 
 = \frac{2\pi}{\rho_s(\sigma_1')}
    \delta(\sigma_{1'}'-\sigma_1)\,4\pi\delta(\Omega_{1'}'-\Omega_{1})$ 
into Eq.~\eqref{eq:unit-ral},
the partial wave unitarity relation is 
\begin{widetext}
\begin{align}
\label{eq:unita-two}
  T(\sigma'_+,s_+,\sigma_+)-T(\sigma'_-,s_+,\sigma_-) = & {\left[T\left(\sigma_{+}^{\prime}, s_{+}, \sigma_{+}\right)-T\left(\sigma_{-}^{\prime}, s_{+}, \sigma_{+}\right)\right]} +\left[T\left(\sigma_{-}^{\prime}, s_{+}, \sigma_{+}\right)-T\left(\sigma_{-}^{\prime}, s_{+}, \sigma_{-}\right)\right] \nonumber \\
  =&\, i\, t^\dagger(\sigma'_+)\rho(\sigma') T(\sigma'_+,s_+,\sigma_+) +
     2i \frac{t^\dagger(\sigma'_+)}{\lambda_{s}^{1/2}(\sigma')} \frac{1}{8\pi}\int_{\sigma^-(\sigma',s_+)}^{\sigma^+(\sigma',s_+)} d\sigma_3' T(\sigma_3',s_+,\sigma_+) \nonumber \\
     &+ i\, T^\dagger(\sigma'_+,s_+,\sigma_+) \rho(\sigma) t(\sigma_+) +
     2i\, \frac{t(\sigma_+)}{\lambda_{s}^{1/2}(\sigma)} \frac{1}{8\pi}\int_{\sigma^-(\sigma,s)}^{\sigma^+(\sigma,s)} d\sigma_3 T^\dagger(\sigma'_+,s_+,\sigma_3) \nonumber \\ 
     &+ 6i\,\frac{2\pi s\, t^\dagger(\sigma'_+)t(\sigma_+)}{\lambda_{s}^{1/2}(\sigma')\lambda_{s}^{1/2}(\sigma)} \,\theta^+(\phi(\sigma',s,\sigma)).
\end{align}
Here, for a real kinematic variable $x$, the subscripts $x_\pm\equiv x\pm i\epsilon$
denote boundary values reached from the upper and lower edges, respectively,
of the relevant unitarity cut, i.e., $s_\pm=s\pm i\epsilon$ refer to the two
boundary values across the three-body cut, whereas
$\sigma_\pm^{(\prime)}=\sigma^{(\prime)} \pm i\epsilon$ refer to those across the corresponding two-body subenergy cuts. In the above equation, the difference on the left-hand side is evaluated sequentially: at fixed
$s_+$, $\sigma'$ is continued from $\sigma'_+$ to $\sigma'_-$ first, and
$\sigma$ is then continued from $\sigma_+$ to $\sigma_-$.
The first two terms are discontinuities across the cuts in the final-state two-particle subenergies $\sigma^{\prime}$ as shown in Fig.~\ref{fig:dis-gro}(a) and Fig.~\ref{fig:dis-gro}(b), the third and fourth terms are discontinuities across the cuts in the initial-state two-particle subenergy $\sigma$ in Fig.~\ref{fig:dis-gro}(c) and Fig.~\ref{fig:dis-gro}(d), and the last term is the discontinuity from the
cross disconnected-disconnected channel in Fig.~\ref{fig:dis-gro}(g).
The two-body phase space $\rho(\sigma)$ is 
\begin{align}
  &\rho(\sigma)\theta(\sigma-4m_{\pi}^2) = \int \frac{d^3 p_2''}{2E_2''(2\pi)^3} \frac{d^3 p_3''}{2E_3''(2\pi)^3} (2\pi)^4\delta^4(q_1''-q_1) = \frac{1}{8\pi} \frac{2|\vec p_2^{\,*}|}{\sqrt{\sigma}}\theta(\sigma-4m_{\pi}^2),
\end{align}
which is evaluated at the c.m. frame of the pair $(23)$ with $|\vec p_2^{\,*}|$ being the break-up momentum $|\vec p^{\,*}| = \frac12\sqrt{\sigma-4m_\pi^2} = \sqrt{\sigma/4-m_\pi^2}$. The Dalitz-plot endpoints are $\sigma^{\pm}(\sigma,s)=G(\sigma,s)\pm F(\sigma,s)$, where
$G(\sigma,s)=\frac{1}{2}(s+3m_{\pi}^2-\sigma)$, $F(\sigma,s)=\lambda_s^{1/2}(\sigma)\lambda^{1/2}(\sigma)/(2\sigma)$.

\section{Dispersive representation}
\label{sec:disrep-full}
Motivated by the Khuri-Treiman equation for the $1\to 3$ decays~\cite{Khuri:1960zz}, we construct the $3\to 3$ scattering amplitude from dispersion relations, using a subenergy as the dispersive integration variable.
As discussed in Refs.~\cite{Aitchison:1966lpz,Pasquier:1968zz}, for the $1\to 3$ decay process the inhomogeneous term guarantees that the amplitude satisfies three-body unitarity, and three-body unitarity is generated automatically from two-body unitarity, analyticity, and crossing symmetry. We will show that the $3\to 3$ scattering amplitude obtained from dispersion relations, using the subenergy as the dispersive integration variable, satisfies three-body unitarity.

Choosing the final-state subenergy variable $\sigma'$ as the dispersive integration
variable at fixed $s_+$ and $\sigma_+$, the first, second and last terms in Eq.~\eqref{eq:unita-two} provide the discontinuity, and thus the dispersive representation of $T(\sigma',s,\sigma_+)$ can be written as
\begin{align}
\label{eq:disprep-ful}
  T(\sigma'_+,s_+,\sigma_+) =& \int_{4m_{\pi}^2}^{+\infty}\frac{d\sigma''}{\sigma''-\sigma'-i\epsilon}\,\frac{6 s\, t^\dagger(\sigma''_+)t(\sigma_+)}{\lambda_{s}^{1/2}(\sigma'')\lambda_{s}^{1/2}(\sigma)} \,\theta^+(\phi(\sigma'',s,\sigma)) + \frac{1}{2\pi}\int_{4m_{\pi}^2}^{+\infty}\frac{d\sigma''} {\sigma''-\sigma'-i\epsilon}t^\dagger(\sigma''_+)\rho(\sigma'') T(\sigma''_+,s_+,\sigma_+)\nonumber \\
 & +\frac{1}{\pi}\int_{4m_{\pi}^2}^{+\infty}\frac{d\sigma''} {\sigma''-\sigma'-i\epsilon}\frac{t^\dagger(\sigma''_+)}{\lambda_{s}^{1/2}(\sigma'')} \frac{1}{8\pi} \int_{\sigma^-(\sigma'',s)}^{\sigma^+(\sigma'',s)} d\sigma_3 \, T(\sigma_{3},s_+,\sigma_+) . 
\end{align}
The definition of $\lambda_s^{1/2}(\sigma)= \lambda^{1/2}(s,\sigma,m_{\pi}^2)$ in the complex plane is given in Ref.~\cite{Bronzan:1963mby}. Following the branch-cut convention discussed in Ref.~\cite{Bronzan:1963mby},
$\lambda_s^{1/2}(\sigma)$ has branch points at $\sigma=(\sqrt{s}\pm m_\pi)^2$,
and is chosen such that
\begin{align}
  \lambda_s^{1/2}(\sigma\pm i\epsilon) &\ge 0,
  && 4m_\pi^2 \le \sigma < (\sqrt{s}-m_\pi)^2, \nonumber \\
  \lambda_s^{1/2}(\sigma+i\epsilon) &= -\,i\,|\lambda_s^{1/2}(\sigma)|,
  && (\sqrt{s}-m_\pi)^2 \le \sigma < (\sqrt{s}+m_\pi)^2,\nonumber \\
  \lambda_s^{1/2}(\sigma-i\epsilon) &= +\,i\,|\lambda_s^{1/2}(\sigma)|,
  && (\sqrt{s}-m_\pi)^2 \le \sigma < (\sqrt{s}+m_\pi)^2, \nonumber\\
  \lambda_s^{1/2}(\sigma\pm i\epsilon) &\le 0,
  && \sigma \ge (\sqrt{s}+m_\pi)^2.
\end{align}
Moreover, in the $s$ plane
\begin{align}
  \lambda_{s\pm i\epsilon}^{1/2}(\sigma) &\ge 0,
  && 4m_\pi^2 \le \sigma < (\sqrt{s}-m_\pi)^2, \nonumber \\
  \lambda_{s- i\epsilon}^{1/2}(\sigma) &= -\,i\,|\lambda_s^{1/2}(\sigma)|,
  && (\sqrt{s}-m_\pi)^2 \le \sigma < (\sqrt{s}+m_\pi)^2,\nonumber \\
  \lambda_{s+ i\epsilon}^{1/2}(\sigma) &= +\,i\,|\lambda_s^{1/2}(\sigma)|,
  && (\sqrt{s}-m_\pi)^2 \le \sigma < (\sqrt{s}+m_\pi)^2, \nonumber\\
  \lambda_{s\pm i\epsilon}^{1/2}(\sigma) &\le 0,
  && \sigma \ge (\sqrt{s}+m_\pi)^2.
  \end{align}
The function $\lambda^{1/2}(\sigma)= \lambda^{1/2}(\sigma,m_{\pi}^2,m_{\pi}^2)$, which enters the endpoint functions $\sigma_\pm$, is chosen as 
  \begin{align}
  \lambda^{1/2}(\sigma\pm i\epsilon) &\ge 0,
  && 4m_\pi^2 \le \sigma .
  \end{align}

\begin{figure}
   \includegraphics[width=0.4\textwidth]{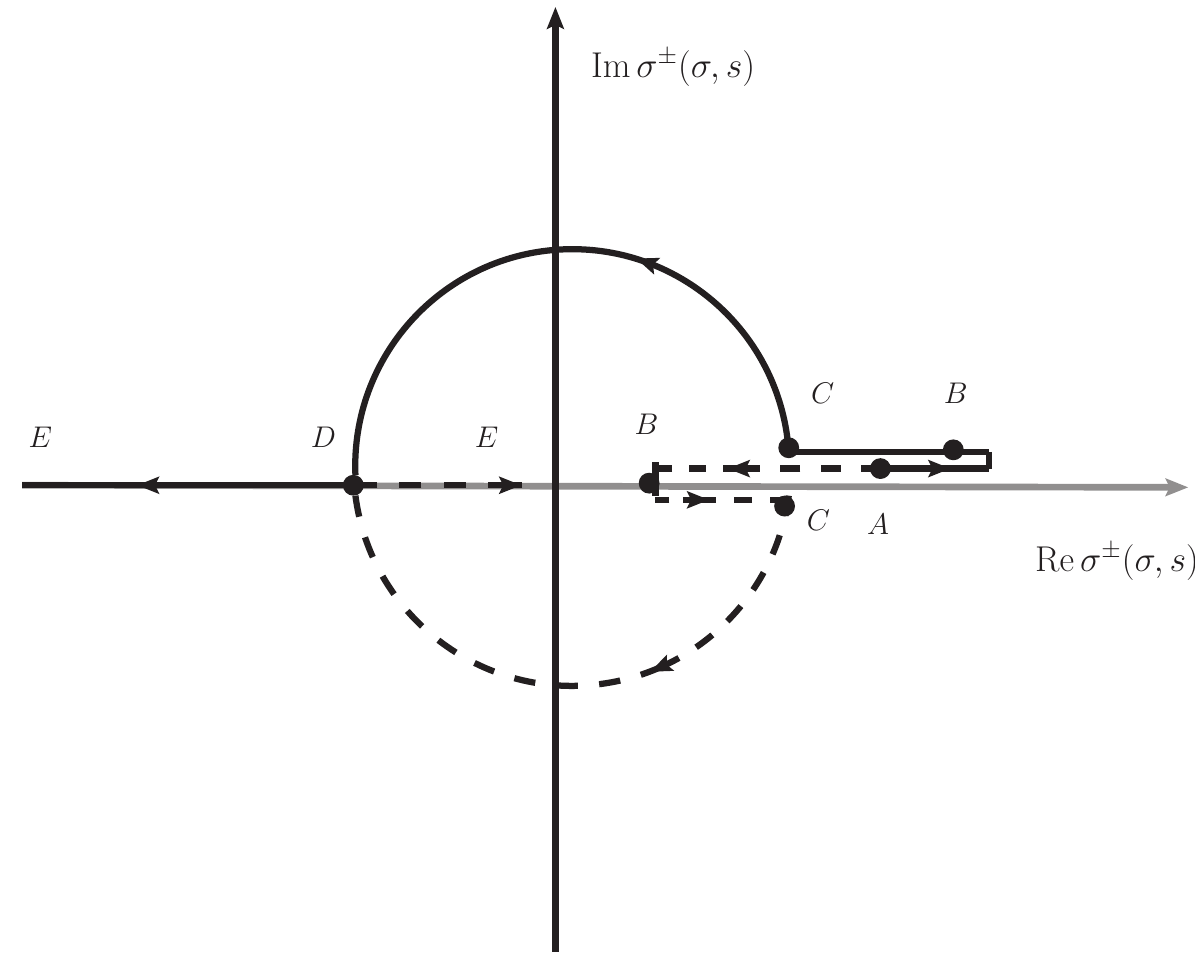}
   \includegraphics[width=0.4\textwidth]{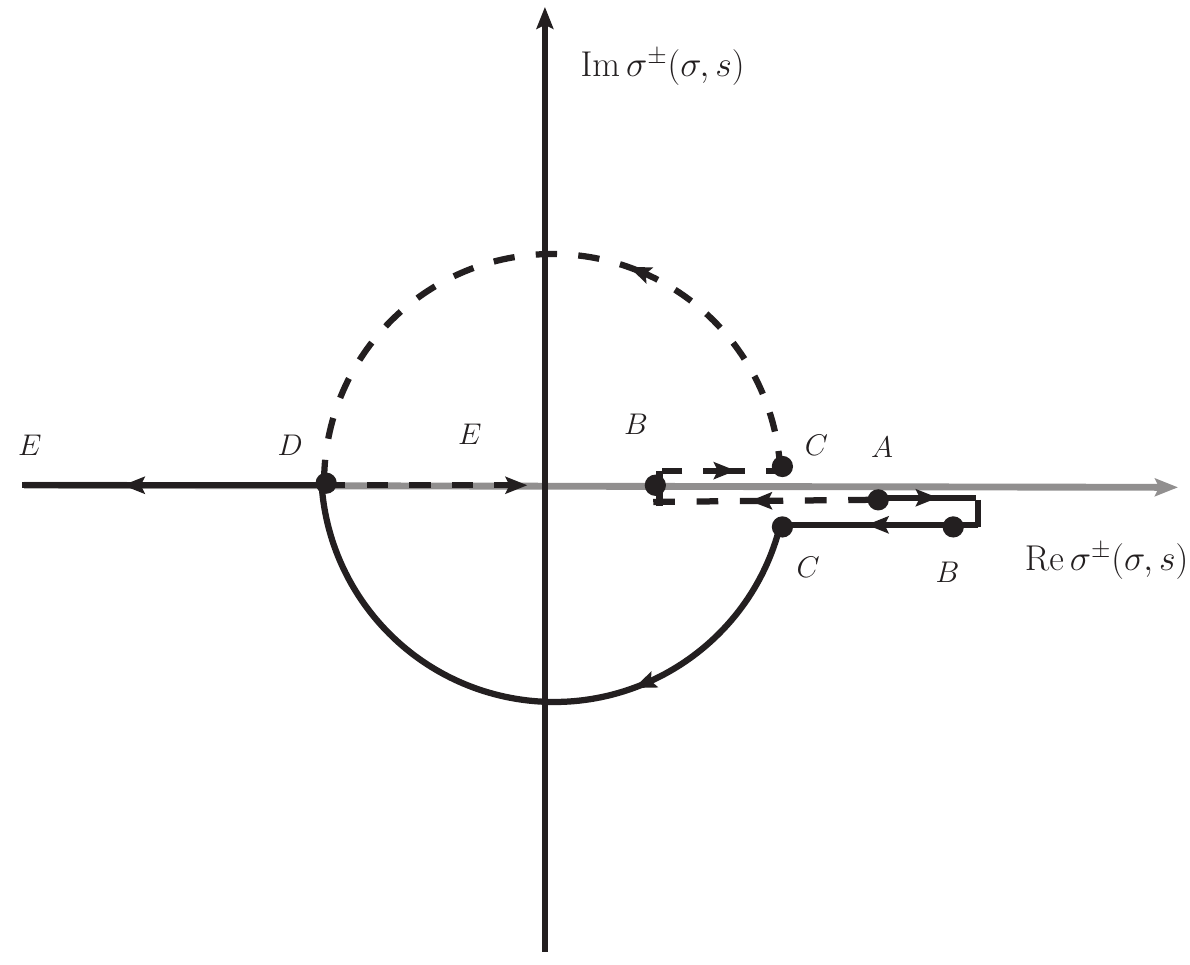}
  \caption{The integration contours for $s+i\epsilon$ and $s-i\epsilon$, respectively. The dashed and solid lines correspond to $\sigma^-(\sigma,s)$ and $\sigma^+(\sigma,s)$, respectively. }
  \label{fig:conto:rea}
\end{figure}

The function $\theta^+(\phi(\sigma'',s,\sigma))$ in the first term in Eq.~\eqref{eq:disprep-ful}
restricts the integration region to the physical Dalitz domain. In particular, for fixed physical $s$ and $\sigma$, the integration domain does not extend outside the region $4m_\pi^2 \le \sigma'' \le (\sqrt{s}-m_\pi)^2$.
Therefore, throughout the integration range, $\lambda_s^{1/2}(\sigma'')$ is real, so the contour never crosses its branch cut. Due to the $\sigma''_+=\sigma''+i\epsilon$ constraint in $t^\dagger(\sigma''_+)$, the integration contour for $t^\dagger(\sigma''_+)$ is taken slightly below $\sigma''$.
The $i\epsilon$ prescription specifies how the Cauchy pole in
$(\sigma''-\sigma'-i\epsilon)^{-1}$ is approached from the physical side; it does not
produce an additional discontinuity across the three-body cut. Consequently, the driving term alone, without being dressed in Eq.~\eqref{eq:disprep-ful}, does not contribute to
the discontinuity across the three-body cut. This is because the driving term does not contain a three-body intermediate state, since at least one of the three particles is external in both the initial and final states.

The three-body unitarity relation of  nonrelativistic scattering has been discussed in Refs.~\cite{Glockle:1983zz,glockle:2011}. In Eq.~\eqref{eq:disprep-ful}, since the first term does not contribute to the discontinuity across the three-body cut, the discontinuity is 
\begin{align}
\label{eq:dis-three}
  T(\sigma'_+,s_+,\sigma_+)-T(\sigma'_+,s_-,\sigma_+)=&\,\frac{1}{2\pi}\int_{4m_{\pi}^2}^{+\infty}\frac{d\sigma''} {\sigma''-\sigma'-i\epsilon}t^\dagger(\sigma''_+)\rho(\sigma'') \Big(T(\sigma''_+,s_+,\sigma_+)-T(\sigma''_+,s_-,\sigma_+)\Big)\nonumber \\
 &\,+\frac{1}{\pi}\int_{4m_{\pi}^2}^{+\infty}\frac{d\sigma''} {\sigma''-\sigma'-i\epsilon} \frac{t^\dagger(\sigma''_+)}{\lambda_{s_+}^{1/2}(\sigma'')} \frac{1}{8\pi} \int_{\sigma^-(\sigma'',s_+)}^{\sigma^+(\sigma'',s_+)} d\sigma_3 \,  T(\sigma_{3},s_+,\sigma_+) \nonumber \\
 &\,-\frac{1}{\pi}\int_{4m_{\pi}^2}^{+\infty}\frac{d\sigma''} {\sigma''-\sigma'-i\epsilon} \frac{t^\dagger(\sigma''_+)}{\lambda_{s_-}^{1/2}(\sigma'')} \frac{1}{8\pi} \int_{\sigma^-(\sigma'',s_-)}^{\sigma^+(\sigma'',s_-)} d\sigma_3 \, T(\sigma_{3},s_-,\sigma_+).
\end{align}
 
For each region of $\sigma$, the integration contour from $\sigma^-(\sigma'',s_\pm)$ to $\sigma^+(\sigma'',s_\pm)$, with the three-body variable $s$ evaluated on the upper and lower edges of the real axis, is shown in the left and right panels of Fig.~\ref{fig:conto:rea}. In the interval $\sigma \in [4m_{\pi}^2,(s-m_\pi^2)/2]$, the upper $\sigma^+(\sigma,s)$ and lower $\sigma^-(\sigma,s)$ integration limits lie in [A,B]; 
in the interval $(s-m_\pi^2)/2 \le \sigma < (\sqrt{s}-m_\pi)^2$, the upper $\sigma^+(\sigma,s)$ and lower $\sigma^-(\sigma,s)$ integration limits lie in [B,C]; 
in the interval $(\sqrt{s}-m_\pi)^2 \le \sigma < (\sqrt{s}+m_\pi)^2$, the upper $\sigma^+(\sigma,s)$ and lower $\sigma^-(\sigma,s)$ integration limits lie in [C,D]; 
in the interval $(\sqrt{s}+m_\pi)^2 \le \sigma$, the upper $\sigma^+(\sigma,s)$ and lower $\sigma^-(\sigma,s)$ integration limits lie in [D,E].

Moreover, with three-body variable $s+i\epsilon$ with an infinitesimal positive imaginary part, 
the endpoints approach the limiting contour as 
  \begin{align}
  \operatorname{Re} \sigma^+(\sigma,s_+)> \operatorname{Re} \sigma^-(\sigma,s_+),&&\operatorname{Im} \sigma^-(\sigma,s_+)> 0, && \operatorname{Im} \sigma^+(\sigma,s_+)> 0, &&
  && 4m_\pi^2 \le \sigma < (s-m_\pi^2)/2, \nonumber \\
   \operatorname{Re} \sigma^+(\sigma,s_+)> \operatorname{Re} \sigma^-(\sigma,s_+),&&\operatorname{Im} \sigma^-(\sigma,s_+)< 0, && \operatorname{Im} \sigma^+(\sigma,s_+)> 0, &&
  && (s-m_\pi^2)/2 < \sigma < (\sqrt{s}-m_\pi)^2, \nonumber \\
   \operatorname{Re} \sigma^+(\sigma,s_+)= \operatorname{Re} \sigma^-(\sigma,s_+),&&\operatorname{Im} \sigma^-(\sigma,s_+)< 0, && \operatorname{Im} \sigma^+(\sigma,s_+)> 0, &&
  &&(\sqrt{s}-m_\pi)^2 \le \sigma < (\sqrt{s}+m_\pi)^2, \nonumber \\
  \operatorname{Re} \sigma^+(\sigma,s_+)< \operatorname{Re} \sigma^-(\sigma,s_+),&&\operatorname{Im} \sigma^-(\sigma,s_+)< 0, && \operatorname{Im} \sigma^+(\sigma,s_+)> 0, &&
  &&(\sqrt{s}+m_\pi)^2 \le \sigma .
  \end{align}
At $\sigma=(s-m_\pi^2)/2$, the $O(\epsilon)$ displacement of $\sigma^-(\sigma,s_\pm)$ vanishes. For $\sigma \geq (\sqrt{s}+m_\pi)^2$, the functions $\sigma^\pm(\sigma,s_+)$ lie on the upper and lower edges of the negative real axis. Since the branch cut is chosen to run from the threshold to the right, no discontinuity arises when crossing the upper and lower edges of the negative real axis. Therefore, the integration contour is taken along the negative real axis. With three-body energy $s-i\epsilon$ with an infinitesimal negative imaginary part, one has
\begin{align}
  \operatorname{Re} \sigma^+(\sigma,s_-)> \operatorname{Re} \sigma^-(\sigma,s_-),&&\operatorname{Im} \sigma^-(\sigma,s_-)< 0, && \operatorname{Im} \sigma^+(\sigma,s_-)< 0, &&
  && 4m_\pi^2 \le \sigma < (s-m_\pi^2)/2, \nonumber \\
   \operatorname{Re} \sigma^+(\sigma,s_-)> \operatorname{Re} \sigma^-(\sigma,s_-),&&\operatorname{Im} \sigma^-(\sigma,s_-)> 0, && \operatorname{Im} \sigma^+(\sigma,s_-)< 0, &&
  && (s-m_\pi^2)/2 < \sigma < (\sqrt{s}-m_\pi)^2, \nonumber \\
   \operatorname{Re} \sigma^+(\sigma,s_-)= \operatorname{Re} \sigma^-(\sigma,s_-),&&\operatorname{Im} \sigma^-(\sigma,s_-)> 0, && \operatorname{Im} \sigma^+(\sigma,s_-)< 0, &&
  &&(\sqrt{s}-m_\pi)^2 \le \sigma < (\sqrt{s}+m_\pi)^2, \nonumber \\
  \operatorname{Re} \sigma^+(\sigma,s_-)< \operatorname{Re} \sigma^-(\sigma,s_-),&&\operatorname{Im} \sigma^-(\sigma,s_-)> 0, && \operatorname{Im} \sigma^+(\sigma,s_-)< 0, &&
  &&(\sqrt{s}+m_\pi)^2 \le \sigma .
\end{align}

We discuss the last two terms in Eq.~\eqref{eq:dis-three}. As $4m_\pi^2 \le \sigma'' < (s-m_\pi^2)/2$, $\lambda_{s_\pm}(\sigma'') >0$ and the integration contours from $\sigma^-(\sigma'',s_+)$ to $\sigma^+(\sigma'',s_+)$ and from $\sigma^-(\sigma'',s_-)$ to $\sigma^+(\sigma'',s_-)$ have the same real part, the only difference is the first one is on the upper edge and second one on the lower edge of the real axis. This is analogous to the region $(s-m_\pi^2)/2 \le \sigma'' < (\sqrt{s}-m_\pi)^2$. For $(\sqrt{s}-m_\pi)^2 \le \sigma'' <
(\sqrt{s}+m_\pi)^2$, the integration contours from $\sigma^-(\sigma'',s_+)$ to $\sigma^+(\sigma'',s_+)$ and from $
\sigma^-(\sigma'',s_-)$ to $\sigma^+(\sigma'',s_-)$ follow the same path, differing only in direction. However, $
\lambda_{s_+}^{1/2}(\sigma'')$ and $\lambda_{s_-}^{1/2}(\sigma'')$ have opposite signs. For $\sigma'' \geq
(\sqrt{s}+m_\pi)^2$, the integration contours from $\sigma^-(\sigma'',s_+)$ to $\sigma^+(\sigma'',s_+)$ and from $
\sigma^-(\sigma'',s_-)$ to $\sigma^+(\sigma'',s_-)$ are identical. Then one has
\begin{align}
  &\,\frac{t^\dagger(\sigma''_+)}{\lambda_{s_+}^{1/2}(\sigma'')} \frac{1}{8\pi} \int_{\sigma^-(\sigma'',s_+)}^{\sigma^+(\sigma'',s_+)} d\sigma_3 \,  T(\sigma_{3},s_+,\sigma_+)-
 \frac{t^\dagger(\sigma''_+)}{\lambda_{s_-}^{1/2}(\sigma'')} \frac{1}{8\pi} \int_{\sigma^-(\sigma'',s_-)}^{\sigma^+(\sigma'',s_-)} d\sigma_3 \, T(\sigma_{3},s_-,\sigma_+)\nonumber \\ 
 =&
\begin{cases}
\displaystyle
\frac{t^\dagger(\sigma''_+)}{\lambda_{s_+}^{1/2}(\sigma'')}
\frac{1}{8\pi}
\int_{\sigma^-(\sigma'',s_+)}^{\sigma^+(\sigma'',s_+)}
d\sigma_3\,
\Bigl[T(\sigma_3,s_+,\sigma_+)
-T(\sigma_3^*,s_-,\sigma_+)\Bigr],
& 4m_\pi^2 \le \sigma'' < (\sqrt{s}-m_\pi)^2,
\\[1.2ex]
\displaystyle
\frac{t^\dagger(\sigma''_+)}{\lambda_{s_+}^{1/2}(\sigma'')}
\frac{1}{8\pi}
\int_{\sigma^-(\sigma'',s_+)}^{\sigma^+(\sigma'',s_+)}
d\sigma_3\,
\Bigl[T(\sigma_3,s_+,\sigma_+)
-T(\sigma_3,s_-,\sigma_+)\Bigr],
& (\sqrt{s}-m_\pi)^2 \le \sigma'' .
\end{cases}
\end{align}

Collecting all, Eq.~\eqref{eq:dis-three} can be written as 
\begin{align}
\label{eq:disc-interm}
T(\sigma'_+,s_+,\sigma_+)-T(\sigma'_+,s_-,\sigma_+)=\frac{1}{2\pi}\int_{4m_{\pi}^2}^{+\infty}\frac{d\sigma''} {\sigma''-\sigma'-i\epsilon}t^\dagger(\sigma''_+)\rho(\sigma'') \Big(T(\sigma''_+,s_+,\sigma_+)-T(\sigma''_+,s_-,\sigma_+)\Big)\nonumber \\
 +\frac{1}{\pi}\int_{4m_{\pi}^2}^{(\sqrt{s}-m_\pi)^2}\frac{d\sigma''} {\sigma''-\sigma'-i\epsilon} \frac{t^\dagger(\sigma''_+)}{\lambda_{s_+}^{1/2}(\sigma'')} \frac{1}{8\pi} \int_{\sigma^-(\sigma'',s_+)}^{\sigma^+(\sigma'',s_+)} d\sigma_3 \, \Big( T(\sigma_{3},s_+,\sigma_+) -T(\sigma_{3}^*,s_-,\sigma_+)\Big)\nonumber \\
 +\frac{1}{\pi}\int_{(\sqrt{s}-m_\pi)^2}^{+\infty}\frac{d\sigma''} {\sigma''-\sigma'-i\epsilon} \frac{t^\dagger(\sigma''_+)}{\lambda_{s_+}^{1/2}(\sigma'')} \frac{1}{8\pi} \int_{\sigma^-(\sigma'',s_+)}^{\sigma^+(\sigma'',s_+)} d\sigma_3 \, \Big( T(\sigma_{3},s_+,\sigma_+) -T(\sigma_{3},s_-,\sigma_+)\Big).
\end{align}
Considering the relation,
\begin{align}
\Big( T(\sigma_{3},s_+,\sigma_+) -T(\sigma_{3}^*,s_-,\sigma_+)\Big)=\Big( T(\sigma_{3},s_-,\sigma_+) -T(\sigma_{3}^*,s_-,\sigma_+)\Big)+\Big( T(\sigma_{3},s_+,\sigma_+) -T(\sigma_{3},s_-,\sigma_+)\Big),
\end{align}
and defining $\Delta T(\sigma_{3},s_+,\sigma_+)\equiv \Big( T(\sigma_{3},s_+,\sigma_+) -T(\sigma_{3},s_-,\sigma_+)\Big)$, we get
\begin{align}
\label{eq:disc-repr}
  \Delta T(\sigma'_+,s_+,\sigma_+)=&\,\frac{1}{2\pi}\int_{4m_{\pi}^2}^{+\infty}\frac{d\sigma''} {\sigma''-\sigma'-i\epsilon}t^\dagger(\sigma''_+)\rho(\sigma'') \Delta T(\sigma''_+,s_+,\sigma_+)\nonumber \\
  &+\frac{1}{\pi}\int_{4m_{\pi}^2}^{+\infty}\frac{d\sigma''} {\sigma''-\sigma'-i\epsilon}\frac{t^\dagger(\sigma''_+)}{\lambda_{s_+}^{1/2}(\sigma'')} \frac{1}{8\pi} \int_{\sigma^-(\sigma'',s_+)}^{\sigma^+(\sigma'',s_+)} d\sigma_3 \,\Delta T(\sigma_3,s_+,\sigma_+)\nonumber \\
&+\frac{1}{\pi}\int_{4m_{\pi}^2}^{(\sqrt{s}-m_\pi)^2}\frac{d\sigma''} {\sigma''-\sigma'-i\epsilon}\frac{t^\dagger(\sigma''_+)}{\lambda_{s_+}^{1/2}(\sigma'')} \frac{1}{8\pi} \int_{\sigma^-(\sigma'',s_+)}^{\sigma^+(\sigma'',s_+)} d\sigma_3 \Big(  T(\sigma_{3},s_-,\sigma_+) -T(\sigma_{3}^*,s_-,\sigma_+) \Big),
\end{align}
where the terms in parentheses in the last line correspond to the discontinuity across the cut in the two-particle subenergy $\sigma_3$ appearing in Eq.~\eqref{eq:unita-two}.
The kernel of $\Delta T(\sigma'_+,s_+,\sigma_+)$ in Eq.~\eqref{eq:disc-repr} is the same as that of $T(\sigma'_+,s_+,\sigma_+)$ in Eq.~\eqref{eq:disprep-ful}.

The driving term of $\Delta T(\sigma'_+,s_+,\sigma_+)$ in Eq.~\eqref{eq:disc-repr} can be written as 
\begin{align}
\label{eq:drive-refor}
&\,\frac{i}{\pi}\int_{4m_{\pi}^2}^{(\sqrt{s}-m_\pi)^2}\frac{d\sigma''} {\sigma''-\sigma'-i\epsilon}\frac{t^\dagger(\sigma''_+)}{\lambda_{s_+}^{1/2}(\sigma'')} \frac{1}{8\pi} \int_{\sigma^-(\sigma'',s_+)}^{\sigma^+(\sigma'',s_+)} d\sigma_3 \Big( \, t^\dagger(\sigma_3)\rho(\sigma_3)
T(\sigma_3,s_-,\sigma_+)  \nonumber \\
&+2 \frac{t^\dagger(\sigma_3)}{\lambda_{s_-}^{1/2}(\sigma_3)} \frac{1}{8\pi}\int_{\sigma^-(\sigma_3,s_-)}^{\sigma^+(\sigma_3,s_-)} d\sigma_3' T(\sigma_3',s_-,\sigma_+)\Big) \nonumber \\
=&\,\frac{i}{\pi}\int_{4m_{\pi}^2}^{(\sqrt{s}-m_{\pi})^2}\frac{d\sigma''} {\sigma''-\sigma'-i\epsilon}\frac{t^\dagger(\sigma''_+)}{\lambda_{s_+}^{1/2}(\sigma'')} \frac{1}{8\pi} \int_{4m_{\pi}^2}^{(\sqrt{s}-m_{\pi})^2} d\sigma_3\theta^+(\phi(\sigma'',s_+,\sigma_{3}))  \Big( \, t^\dagger(\sigma_{3})\rho(\sigma_3)
T(\sigma_{3},s_-,\sigma_+)  \nonumber \\
&\,+2 \frac{t^\dagger(\sigma_3)}{\lambda_{s_-}^{1/2}(\sigma_3)} \frac{1}{8\pi}\int_{\sigma^-(\sigma_3,s_-)}^{\sigma^+(\sigma_3,s_-)} d\sigma_3' T(\sigma_3',s_-,\sigma_+)\Big)\nonumber \\
=&\,\frac{i}{6\pi}\int_{4m_{\pi}^2}^{(\sqrt{s}-m_{\pi})^2}\frac{d\sigma''} {\sigma''-\sigma'-i\epsilon}\frac{6st^\dagger(\sigma''_+)}{\lambda_{s_+}^{1/2}(\sigma'')} \int_{4m_{\pi}^2}^{(\sqrt{s}-m_{\pi})^2} d\sigma_3   \frac{t(\sigma_{3})}{\lambda_{s_+}^{1/2}(\sigma_3)} \theta^+(\phi(\sigma'',s_+,\sigma_{3}))  \rho(\sigma_3) \rho_{s}(\sigma_3)
T(\sigma_{3}^*,s_-,\sigma_+)\nonumber  \\
&\,+\frac{i}{3\pi s}\frac{1}{(8\pi)^2}\int_{4m_{\pi}^2}^{(\sqrt{s}-m_{\pi})^2}\frac{d\sigma''} {\sigma''-\sigma'-i\epsilon}\frac{6st^\dagger(\sigma''_+)}{\lambda_{s_-}^{1/2}(\sigma'')} \int_{4m_{\pi}^2}^{(\sqrt{s}-m_{\pi})^2} d\sigma_3   \frac{t(\sigma_{3})}{\lambda_{s_-}^{1/2}(\sigma_3)} \theta^+(\phi(\sigma'',s_+,\sigma_{3})) \int_{\sigma^-(\sigma_3,s_-)}^{\sigma^+(\sigma_3,s_-)} d\sigma_3' T(\sigma_3'^*,s_-,\sigma_+),
\end{align}
where we have used the relations
\begin{align}
\label{eq:comp-rela}
t^\dagger(\sigma_{3})\rho(\sigma_3)
T(\sigma_{3},s_-,\sigma_+) &=t(\sigma_{3}) \rho(\sigma_3) 
T(\sigma_{3}^*,s_-,\sigma_+),  \nonumber \\
\frac{t^\dagger(\sigma_3)}{\lambda_{s_-}^{1/2}(\sigma_3)} \int_{\sigma^-(\sigma_3,s_-)}^{\sigma^+(\sigma_3,s_-)} d\sigma_3' T(\sigma_3',s_-,\sigma_+)
&= \frac{t(\sigma_{3})}{\lambda_{s_-}^{1/2}(\sigma_3)} \int_{\sigma^-(\sigma_3,s_-)}^{\sigma^+(\sigma_3,s_-)} d\sigma_3' T(\sigma_3'^*,s_-,\sigma_+).
\end{align}
The driving term in Eq.~\eqref{eq:drive-refor} is a linear combination of the driving terms in Eq.~\eqref{eq:disprep-ful}. Note that, for real $\sigma$, the integration region of the driving term in Eq.~\eqref{eq:disprep-ful} is
restricted to the interval from $4m_{\pi}^2$ to $(\sqrt{s}-m_{\pi})^2$ by the constraint imposed by $\theta^+(\phi(\sigma'',s,\sigma))$.
All together, we can conclude that the solution must be a linear combination of  the solution of Eq.~\eqref{eq:disprep-ful}, and can be written as 
\begin{align} 
  T(\sigma'_+,s_+,\sigma_+)-T(\sigma'_+,s_-,\sigma_+) =&\,
  \frac{i}{6\pi}\,\int_{4m_{\pi}^2}^{(\sqrt{s}-m_{\pi})^2} d\sigma''\, T(\sigma'_+,s_+,\sigma'') \rho(\sigma'')\rho_s(\sigma'') T(\sigma''^*,s_-,\sigma_+)\nonumber \\
  &+\frac{i}{3\pi s}\,\frac{1}{(8\pi)^2}\iint_{\phi(\sigma_3,s,\sigma_3')>0} {d \sigma_3d\sigma_3'}\, T(\sigma'_+,s_+,\sigma_{3}) T(\sigma_{3}'^*,s_-,\sigma_+).
\end{align}
This has the same form as that derived in Refs.~\cite{Fleming:1964zz,Mikhasenko:2019vhk} for the three-body discontinuity across the three-body cut. We conclude that the amplitude in Eq.~\eqref{eq:disprep-ful} satisfies three-body unitarity, and three-body unitarity is generated automatically from two-body unitarity, analyticity, and crossing symmetry.

\section{Pair-wise interaction}
\label{sec:disrep-pair}

The construction for the $3\to 3 $ scattering amplitude can be simplified somewhat if the two-particle scattering amplitude can be factored out of $T(\sigma',s,\sigma)$, 
\begin{align} 
\label{eq:pair-dico}
  T(\sigma',s,\sigma)=t(\sigma') M(\sigma',s,\sigma)t(\sigma).
\end{align}
Inserting Eq.~\eqref{eq:pair-dico} into Eq.~\eqref{eq:unita-two}, one obtains the partial-wave unitarity relation for $ M(\sigma'_+,s_+,\sigma_+)$, 
\begin{align}
\label{eq:unita-two-pair}
  & M(\sigma'_+,s_+,\sigma_+)-M(\sigma'_-,s_+,\sigma_-) =6i\,\frac{2\pi s\, }{\lambda_s^{1/2}(\sigma')\lambda_s^{1/2}(\sigma)} \,\theta^+(\phi(\sigma',s,\sigma)) \nonumber \\
  &  +\frac{2i}{\lambda_s^{1/2}(\sigma')} \frac{1}{8\pi}\int_{\sigma^-(\sigma',s)}^{\sigma^+(\sigma',s)} d\sigma_3't(\sigma_3') M(\sigma_3',s_+,\sigma_+) 
  +
  \frac{2i}{\lambda_s^{1/2}(\sigma)} \frac{1}{8\pi}\int_{\sigma^-(\sigma,s)}^{\sigma^+(\sigma,s)} d\sigma_3 t^\dagger(\sigma_3)M^\dagger(\sigma'_+,s_+,\sigma_3).
 \end{align}
The dispersive representation of $M(\sigma'_+,s_+,\sigma_+)$ can be written as
\begin{align}
\label{eq:disrep-pair}
 & M(\sigma'_+,s_+,\sigma_+) =\int_{4m_{\pi}^2}^{+\infty} d\sigma''\frac{1}{\sigma''-\sigma'-i\epsilon}\,\frac{6 s\, }{\lambda_s^{1/2}(\sigma'')\lambda_s^{1/2}(\sigma)} \,\theta^+(\phi(\sigma'',s,\sigma)) \nonumber \\
 &  +\frac{1}{\pi}\int_{4m_{\pi}^2}^{+\infty} d\sigma''\frac{1}{\sigma''-\sigma'-i\epsilon}\frac{1}{\lambda_s^{1/2}(\sigma'')} \frac{1}{8\pi}\int_{4m_{\pi}^2}^{(\sqrt{s}-m_{\pi})^2} d\sigma_3' \,\theta^+(\phi(\sigma'',s,\sigma_3'))t(\sigma_3') M(\sigma_3',s,\sigma).
\end{align}
For the first term, the endpoint contour and the branch of $\lambda_s^{1/2}(\sigma'')$ are continued from the physical region to $s_+$, with the branch convention specified above.
It gives 
\begin{align}
&\int_{4m_{\pi}^2}^{+\infty} d\sigma''\frac{1}{\sigma''-\sigma'-i\epsilon}\,\frac{6 s\, }{\lambda_s^{1/2}(\sigma'')\lambda_s^{1/2}(\sigma)} \,\theta^+(\phi(\sigma'',s,\sigma)) \nonumber \\
&=\int_{\sigma^-(\sigma,s_+)}^{\sigma^+(\sigma,s_+)} d\sigma''\frac{1}{\sigma''-\sigma'-i\epsilon}\,\frac{6 s\, }{\lambda_s^{1/2}(\sigma'')\lambda_s^{1/2}(\sigma)} \,\nonumber \\
&= 
\frac{6 s\, }{\lambda_s^{1/2}(\sigma')\lambda_s^{1/2}(\sigma)}{\text{log}}\Big(\frac{-s^2+s(\sigma'+\sigma)+(\sigma'-m_{\pi}^2)(\sigma-m_{\pi}^2)-\lambda_s^{1/2}(\sigma')\lambda_s^{1/2}(\sigma)}{-s^2+s(\sigma'+\sigma)+(\sigma'-m_{\pi}^2)(\sigma-m_{\pi}^2)+\lambda_s^{1/2}(\sigma')\lambda_s^{1/2}(\sigma)}  \Big),
\end{align}
which is identified with the partial-wave one-particle-exchange potential in Ref.~\cite{Jackura:2018xnx}.

As discussed in Section~\ref{sec:disrep-full}, the driving term in Eq.~\eqref{eq:disrep-pair} does not contribute to the discontinuity across the three-body cut. The discontinuity across the three-body cut is
\begin{align}
\label{eq:disc-pair-thr}
 M(\sigma'_+,s_+,\sigma_+) -M(\sigma'_+,s_-,\sigma_+)=&\, \frac{1}{\pi}\int_{4m_{\pi}^2}^{+\infty} d\sigma''\frac{1}{\sigma''-\sigma'-i\epsilon}\frac{1}{\lambda_{s_+}^{1/2}(\sigma'')} \frac{1}{8\pi}\int_{\sigma^-(\sigma'',s_+)}^{\sigma^+(\sigma'',s_+)} d\sigma_3't(\sigma_3') M(\sigma_3',s_+,\sigma_+) \nonumber \\
 & -\frac{1}{\pi}\int_{4m_{\pi}^2}^{+\infty} d\sigma''\frac{1}{\sigma''-\sigma'-i\epsilon}\frac{1}{\lambda_{s_-}^{1/2}(\sigma'')} \frac{1}{8\pi}\int_{\sigma^-(\sigma'',s_-)}^{\sigma^+(\sigma'',s_-)} d\sigma_3't(\sigma_3') M(\sigma_3',s_-,\sigma_+).
\end{align}
Analogously to Eq.~\eqref{eq:disc-interm}, Eq.~\eqref{eq:disc-pair-thr} can be written as 
\begin{align}
 & M(\sigma'_+,s_+,\sigma_+) -M(\sigma'_+,s_-,\sigma_+) \nonumber \\
=&\,\frac{1}{\pi}\int_{4m_{\pi}^2}^{(\sqrt{s}-m_{\pi})^2} d\sigma''\frac{1}{\sigma''-\sigma'-i\epsilon}\frac{1}{\lambda_{s_+}^{1/2}(\sigma'')} \frac{1}{8\pi}\int_{\sigma^-(\sigma'',s_+)}^{\sigma^+(\sigma'',s_+)} d\sigma_3\Big(t(\sigma_3) M(\sigma_3,s_+,\sigma_+)-t(\sigma_3^*) M(\sigma_3^*,s_-,\sigma_+)   \Big) \nonumber \\
&+\frac{1}{\pi}\int_{(\sqrt{s}-m_{\pi})^2}^{+\infty} d\sigma''\frac{1}{\sigma''-\sigma'-i\epsilon}\frac{1}{\lambda_{s_+}^{1/2}(\sigma'')} \frac{1}{8\pi}\int_{\sigma^-(\sigma'',s_+)}^{\sigma^+(\sigma'',s_+)} d\sigma_3\Big(t(\sigma_3) M(\sigma_3,s_+,\sigma_+)-t(\sigma_3) M(\sigma_3,s_-,\sigma_+)   \Big). \nonumber \\
\end{align}
Considering the relation,
\begin{align}
&\Big( t(\sigma_{3})M(\sigma_{3},s_+,\sigma_+) -t(\sigma_{3}^*)M(\sigma_{3}^*,s_-,\sigma_+)\Big)\nonumber\\
=&\,\Big(t(\sigma_{3}) M(\sigma_{3},s_-,\sigma_+) -t(\sigma_{3}^*)M(\sigma_{3}^*,s_-,\sigma_+)\Big)+\Big( t(\sigma_{3})M(\sigma_{3},s_+,\sigma_+) -t(\sigma_{3})M(\sigma_{3},s_-,\sigma_+)\Big)\,,
\end{align}
and defining $t(\sigma_{3})\Delta M(\sigma_{3},s_+,\sigma_+)\equiv \Big(t(\sigma_{3}) M(\sigma_{3},s_+,\sigma_+) -t(\sigma_{3})M(\sigma_{3},s_-,\sigma_+)\Big)$, we can get 
\begin{align}
\label{eq:disc-repr-pair}
&\Delta M(\sigma'_+,s_+,\sigma_+)\nonumber \\
=&\,\frac{1}{\pi}\int_{4m_{\pi}^2}^{+\infty}\frac{d\sigma''} {\sigma''-\sigma'-i\epsilon} \frac{1}{\lambda_{s_+}^{1/2}(\sigma'')} \frac{1}{8\pi} \int_{\sigma^-(\sigma'',s_+)}^{\sigma^+(\sigma'',s_+)} d\sigma_3 \,t(\sigma_3) \Delta M(\sigma_3,s_+,\sigma_+)\nonumber \\
&+\frac{1}{\pi}\int_{4m_{\pi}^2}^{(\sqrt{s}-m_{\pi})^2}\frac{d\sigma''} {\sigma''-\sigma'-i\epsilon} \frac{1}{\lambda_{s_+}^{1/2}(\sigma'')} \frac{1}{8\pi} \int_{\sigma^-(\sigma'',s_+)}^{\sigma^+(\sigma'',s_+)} d\sigma_3 \,\Big(t(\sigma_{3}) M(\sigma_{3},s_-,\sigma_+) -t(\sigma_{3}^*)M(\sigma_{3}^*,s_-,\sigma_+)\Big).
\end{align}
The kernel of $\Delta M(\sigma'_+,s_+,\sigma_+)$ in Eq.~\eqref{eq:disc-repr-pair} is the same as that of $M(\sigma'_+,s_+,\sigma_+)$ in Eq.~\eqref{eq:disrep-pair}. Using the discontinuities across the cuts in the two-body subenergy $\sigma_3$ appearing in Eq.~\eqref{eq:unita-two-pair}, the last term in Eq.~\eqref{eq:disc-repr-pair} can be written as 
\begin{align}
  \label{eq:drive-refor-pair}
  &\,\frac{i}{\pi}\int_{4m_{\pi}^2}^{(\sqrt{s}-m_\pi)^2}\frac{d\sigma''} {\sigma''-\sigma'-i\epsilon}\frac{1}{\lambda_{s_+}^{1/2}(\sigma'')} \frac{1}{8\pi} \int_{\sigma^-(\sigma'',s_+)}^{\sigma^+(\sigma'',s_+)} d\sigma_3 \Big( \, t^\dagger(\sigma_3)\rho(\sigma_3)
  t(\sigma_3)M(\sigma_3,s_-,\sigma_+)  \nonumber \\
  &+2 \frac{t^\dagger(\sigma_3)}{\lambda_{s_-}^{1/2}(\sigma_3)} \frac{1}{8\pi}\int_{\sigma^-(\sigma_3,s_-)}^{\sigma^+(\sigma_3,s_-)} d\sigma_3' {t(\sigma_3')}M(\sigma_3',s_-,\sigma_+)\Big) \nonumber \\
  =&\,\frac{i}{\pi}\int_{4m_{\pi}^2}^{(\sqrt{s}-m_{\pi})^2}\frac{d\sigma''} {\sigma''-\sigma'-i\epsilon}\frac{1}{\lambda_{s_+}^{1/2}(\sigma'')} \frac{1}{8\pi} \int_{4m_{\pi}^2}^{(\sqrt{s}-m_{\pi})^2} d\sigma_3\theta^+(\phi(\sigma'',s_+,\sigma_{3}))  \Big( \, t^\dagger(\sigma_{3})\rho(\sigma_3)t(\sigma_3)
  M(\sigma_{3},s_-,\sigma_+)  \nonumber \\
  &+2 \frac{t^\dagger(\sigma_3)}{\lambda_{s_-}^{1/2}(\sigma_3)} \frac{1}{8\pi}\int_{\sigma^-(\sigma_3,s_-)}^{\sigma^+(\sigma_3,s_-)} d\sigma_3' t(\sigma_3')M(\sigma_3',s_-,\sigma_+)\Big)\nonumber \\
  =&\,\frac{i}{6\pi}\int_{4m_{\pi}^2}^{(\sqrt{s}-m_{\pi})^2}\frac{d\sigma''} {\sigma''-\sigma'-i\epsilon}\frac{6s}{\lambda_{s_+}^{1/2}(\sigma'')} \int_{4m_{\pi}^2}^{(\sqrt{s}-m_{\pi})^2} d\sigma_3   \frac{1}{\lambda_{s_+}^{1/2}(\sigma_3)} \theta^+(\phi(\sigma'',s_+,\sigma_{3})) t(\sigma_{3}) \rho(\sigma_3) \rho_{s}(\sigma_3)
  t(\sigma_{3}^*)M(\sigma_{3}^*,s_-,\sigma_+)\nonumber  \\
  &+\frac{i}{3\pi s}\frac{1}{(8\pi)^2}\int_{4m_{\pi}^2}^{(\sqrt{s}-m_{\pi})^2}\frac{d\sigma''} {\sigma''-\sigma'-i\epsilon}\frac{6s}{\lambda_{s_-}^{1/2}(\sigma'')} \int_{4m_{\pi}^2}^{(\sqrt{s}-m_{\pi})^2} d\sigma_3   \frac{1}{\lambda_{s_-}^{1/2}(\sigma_3)} \theta^+(\phi(\sigma'',s_+,\sigma_{3})) t(\sigma_{3}) \int_{\sigma^-(\sigma_3,s_-)}^{\sigma^+(\sigma_3,s_-)} d\sigma_3' \nonumber \\
  &\times t(\sigma_3'^*) M(\sigma_3'^*,s_-,\sigma_+).
\end{align}
Analogously to the discussion in Section~\ref{sec:disrep-full}, 
the discontinuity can be written as
\begin{align} 
  M(\sigma'_+,s_+,\sigma_+)-M(\sigma'_+,s_-,\sigma_+) =&\,
  \frac{i}{6\pi}\,\int_{4m_{\pi}^2}^{(\sqrt{s}-m_{\pi})^2} d\sigma''\, M(\sigma'_+,s_+,\sigma'') t(\sigma'')\rho(\sigma'')\rho_s(\sigma'') t(\sigma''^*)M(\sigma''^*,s_-,\sigma_+)\nonumber \\
  &+\frac{i}{3\pi s}\,\frac{1}{(8\pi)^2}\iint_{\phi(\sigma_3,s,\sigma_3')>0} {d \sigma_3d\sigma_3'}\, M(\sigma'_+,s_+,\sigma_{3})t(\sigma_{3})t(\sigma_{3}'^*) M(\sigma_{3}'^*,s_-,\sigma_+).
\end{align}
We conclude that, similarly to the discussion in Section~\ref{sec:disrep-full}, the amplitude with pair-wise interactions in Eq.~\eqref{eq:disrep-pair} also satisfies three-body unitarity, which is generated automatically from two-body unitarity, analyticity, and crossing symmetry.

\section{Connecting to other parameterizations}
\label{sec:demonst}

In this section, we demonstrate that the parameterization in terms of pair-wise interactions is connected to the parameterizations in Refs.~\cite{Freedman:1966xx,Aaron:1968aoz,Mai:2017vot,Jackura:2018xnx}. On the physical Riemann sheet, $t(\sigma') M(\sigma',s,\sigma)$ is analytic in the $\sigma'$ plane, with a cut along the real axis for $\sigma' \geq 4m_{\pi}^2$. Using Cauchy's theorem and assuming no bound state and convergence,
\begin{align}
t(\sigma') M(\sigma',s,\sigma)=\frac{1}{2i\pi}\oint_{\gamma_2}\frac{d\sigma''}{\sigma''-\sigma'-i\epsilon} t(\sigma'') M(\sigma'',s,\sigma),
\end{align}
where the contour $\gamma_2$ is a clockwise contour encircling the two-particle
unitarity cut $\sigma' \geq 4m_{\pi}^2$. Then one gets 
\begin{align}
\label{eq:conn-demon}
 M(\sigma'_+,s_+,\sigma_+) =& \int_{4m_{\pi}^2}^{+\infty} d\sigma''\frac{1}{\sigma''-\sigma'-i\epsilon}\,\frac{6 s\, }{\lambda_s^{1/2}(\sigma'')\lambda_s^{1/2}(\sigma)} \,\theta^+(\phi(\sigma'',s,\sigma)) \nonumber \\
 &  +\frac{1}{\pi}\int_{4m_{\pi}^2}^{+\infty} d\sigma''\frac{1}{\sigma''-\sigma'-i\epsilon}\frac{1}{\lambda_s^{1/2}(\sigma'')} \frac{1}{8\pi}\int_{\sigma^-(\sigma'',s)}^{\sigma^+(\sigma'',s)}d\sigma_3' \,\frac{1}{2i\pi}\oint_{\gamma_2} \frac{d\sigma_3}{\sigma_3-\sigma_3'-i\epsilon} t(\sigma_3) M(\sigma_3,s,\sigma).
\end{align}
By bringing the $d\sigma_3'$ integral to the front, an equation identical to Eq.~\eqref{eq:conn-demon} is obtained, since the functions in both equations are evaluated at the same well-defined values. The kernel of the equation is 
\begin{align}
 & K(\sigma'_+,s_+,\sigma_{3+}) =\frac{1}{2i\pi^2}\frac{1}{8\pi}\int_{4m_{\pi}^2}^{+\infty} d\sigma''\frac{1}{\sigma''-\sigma'-i\epsilon}\frac{1}{\lambda_s^{1/2}(\sigma'')} \int_{\sigma^-(\sigma'',s)}^{\sigma^+(\sigma'',s)}d\sigma_3' \,\frac{1}{\sigma_3-\sigma_3'-i\epsilon}.
\end{align}
The kernel is identical to the triangle loop diagram, and the associated singularities have been discussed in Refs.~\cite{Barton:1961xw,Aitchison:1965xx,Kacser:1966xx,Bronzan:1963mby}. 
Shrinking the $\gamma_2$ contour, we get 
\begin{align}
  M(\sigma'_+,s_+,\sigma_+) =& \int_{4m_{\pi}^2}^{+\infty} d\sigma''\frac{1}{\sigma''-\sigma'-i\epsilon}\,\frac{6 s\, }{\lambda_s^{1/2}(\sigma'')\lambda_s^{1/2}(\sigma)} \,\theta^+(\phi(\sigma'',s,\sigma)) \nonumber \\
  &  +\int_\Gamma {d\sigma_3} \Delta K(\sigma'_+,s_+,\sigma_{3+})t(\sigma_{3+}) M(\sigma_{3+},s_+,\sigma_+),
\end{align}
where $\Delta K(\sigma'_+,s_+,\sigma_{3+})$ is the discontinuity across the cut in $\sigma_{3+}$,
\begin{align}
 \Delta K(\sigma'_+,s_+,\sigma_{3+}) & =K(\sigma'_+,s_+,\sigma_{3-})-K(\sigma'_+,s_+,\sigma_{3+})\nonumber \\
&= \frac{1}{2i\pi^2}\frac{1}{8\pi}\int_{4m_{\pi}^2}^{+\infty} d\sigma''\frac{1}{\sigma''-\sigma'-i\epsilon}\frac{1}{\lambda_s^{1/2}(\sigma'')} \int_{\sigma^-(\sigma'',s)}^{\sigma^+(\sigma'',s)}d\sigma_3' \,2i\pi \delta(\sigma_3'-\sigma_3)\nonumber \\
&=\frac{1}{\pi}\frac{1}{8\pi}\int_{\sigma^-({\sigma_3},s)}^{\sigma^+({\sigma_3},s)} d\sigma''\frac{1}{\sigma''-\sigma'-i\epsilon}\frac{1}{\lambda_s^{1/2}(\sigma'')}. 
\end{align}
As discussed in Refs.~\cite{Aitchison:1966lpz,Pasquier:1968zz}, the shrinking of the $\gamma_2$ contour is stopped on the upper side by the end-point singularities at $\sigma^\pm$. The resulting contour $\Gamma$ can be placed on the upper side of the real axis from $-\infty$ to $(\sqrt{s}-m_\pi)^2$ and passes over the short cut joining the endpoint singularities
$\sigma^-(\sigma'_+,s_+)$ and $\sigma^+(\sigma'_+,s_+)$ 
on its upper side. 
With the definition $\tau(\sigma)=t(\sigma)\rho_s(\sigma)$, one gets 
\begin{align}
  M(\sigma'_+,s_+,\sigma_+) =6s(8\pi^2)\frac{\Delta K(\sigma'_+,s_+,\sigma_+)}{\lambda_s^{1/2}(\sigma)}
  +8\pi s
  \int_\Gamma 
  d\sigma_3\,\frac{\Delta K(\sigma'_+,s_+,\sigma_{3+})}{\lambda_s^{1/2}(\sigma_{3})}\tau(\sigma_{3+}) M(\sigma_{3+},s_+,\sigma_+).
\end{align} 
This equation has the same form as the parameterizations in Refs.~\cite{Freedman:1966xx,Aaron:1968aoz,Mai:2017vot,Jackura:2018xnx}, where three-body unitarity has been discussed. 
\end{widetext}

\section{Analytic continuation}
\label{sec:continua}

\begin{figure}
   \includegraphics[width=0.46\textwidth]{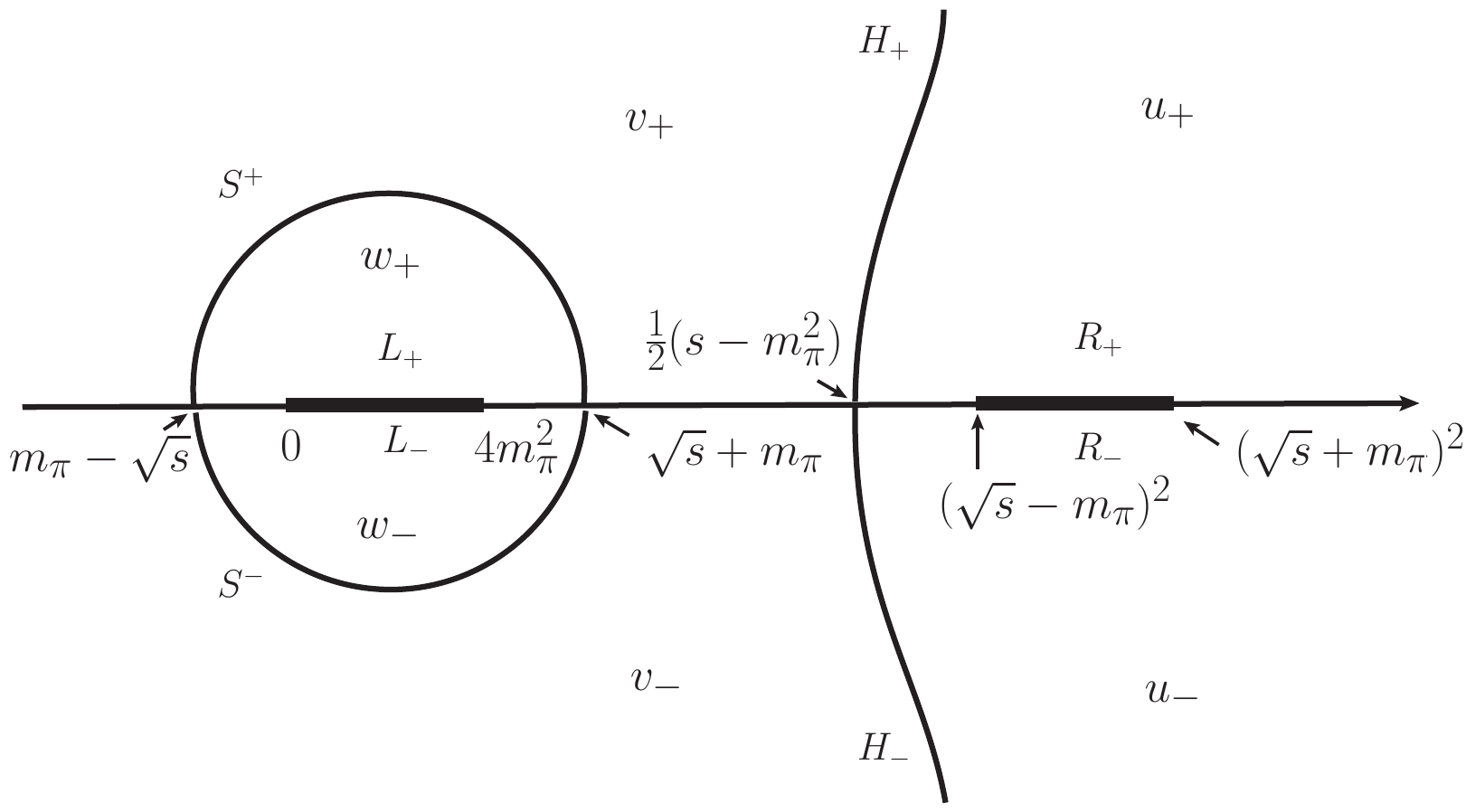}
  \caption{Boundaries of the mapping regions. }
  \label{fig:boundary}
\end{figure}
Once the amplitude is fixed on the real axis, its analytic structure is uniquely determined and can be studied in detail. Three-body unitarity gives rise to a branch cut along the real axis, extending from the three-body threshold to infinity, and thereby generates the structure of the unphysical Riemann sheets. The physical Riemann sheet
associated with the three-body cut is the one that contains the physical values of the amplitude just above the real axis. In the dispersive representation, the amplitude has no singularities on the physical Riemann sheet other than the unitarity cut.

In this section, we discuss the analytic continuation across the three-body cut. By rotating the integration contour, one can analytically continue the three-body amplitude through the unitarity cut onto the unphysical Riemann sheet of the complex energy plane. This continuation may reveal poles of the scattering amplitude that, when located sufficiently
close to the real energy axis, can be identified with resonances. In the following, we implement this contour rotation to extend the energy domain to the unphysical Riemann sheet of the complex energy plane, where such resonance poles may reside.
\begin{figure*}
   \includegraphics[width=0.48\textwidth]{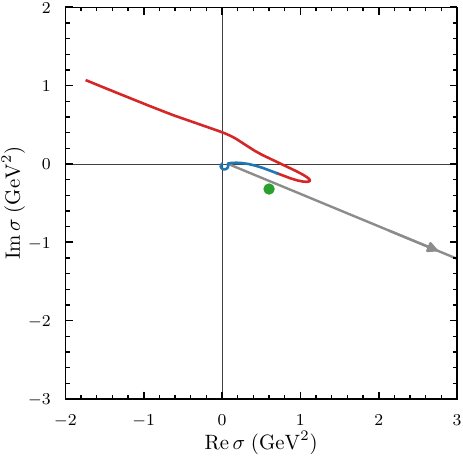}
   \includegraphics[width=0.48\textwidth]{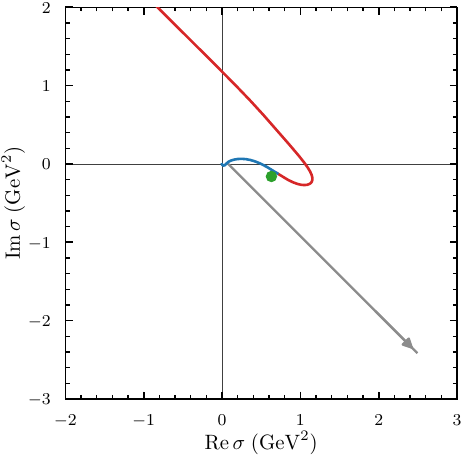}\\[3mm] 
   \includegraphics[width=0.48\textwidth]{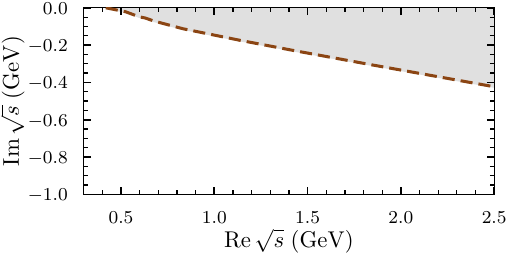}
   \includegraphics[width=0.48\textwidth]{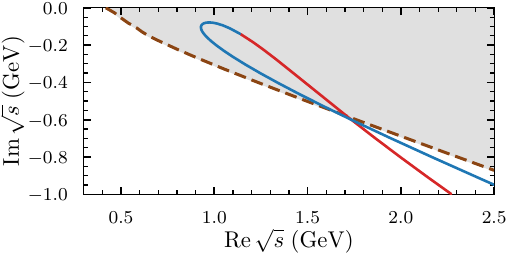}
  \caption{Upper row: The grey lines with arrows are the dispersive integration contours in Eq.~\eqref{eq:disprep-ful}. These contours are rotated by an angle $\phi=\pi/8$ and $\phi=\pi/4$ in the left and right panels, respectively.
  The red and blue lines are the integration contours of ${\sigma^+(\sigma'',s)}$ and ${\sigma^-(\sigma'',s)}$, respectively. We have taken the physical pion mass for $m_\pi$ and $\sqrt{s}=1.2-0.1i$~GeV. The green points represent the pole positions on the unphysical Riemann sheets in the two-body interactions. The pole positions are taken to be $\sqrt{\sigma}=0.8-0.2i$~GeV and $\sqrt{\sigma}=0.8-0.1i$~GeV in the left and right panels, respectively.\\
 Lower row: The domain of analyticity of the unphysical sheet in the lower-half $\sqrt{s}$ plane (grey region). The brown dashed lines mark the boundaries of the analyticity region. The red and blue lines correspond to $\sigma_\mathrm{II}^{\pm}(\sigma'',s)=\sigma_\mathrm{pole}$.}
  \label{fig:reference.plane}
\end{figure*}

We study a crucial mapping transformation and take a first step into the complex plane. The mapping $\sigma\to\sigma^{\pm}(\sigma,s)$ for real $\sigma$ and $s$ has been given in Section~\ref{sec:disrep-full}. In this section, writing $\sigma=x+iy$ with $x,y\in\mathbb{R}$, we discuss the
mapping $\sigma\to\sigma^{\pm}(\sigma,s)$ for complex $\sigma$ and $s$; related treatments can be found in Refs.~\cite{Kacser:1963zz,Bonnevay:1963xx,Aitchison:1964rwb}. The $\sigma$ plane can be divided into a few regions, as shown
in Fig.~\ref{fig:boundary}. The boundary curves can be obtained by determining the complex values of $\sigma$ for which either $\sigma^{+}(\sigma,s)$ or $\sigma^{-}(\sigma,s)$ is real. For $y\neq 0$, since $\operatorname{Im}\,G=-\frac{1}{2}y=\mp \operatorname{Im}\,F$, one has $F=\operatorname{Re}\,F\pm \frac{1}{2}iy$ and, hence, $F^2=[(\operatorname{Re}\,F)^2-
  y^2/4]\pm iy\,\operatorname{Re}\,F$. Therefore,
\begin{align}
  \operatorname{Re}F^2=\left(\frac{\operatorname{Im}F^2}{y}\right)^2-\frac{y^2}{4}.
\end{align}
Using $F(\sigma)=\lambda_s^{1/2}(\sigma)\lambda^{1/2}(\sigma,m_\pi^2,m_\pi^2)/(2\sigma)$, one obtains
\begin{align}
  &y^2= \nonumber \\
  &-\frac{[x-\frac{1}{2}(s-m_{\pi}^2)][x-(m_{\pi}^2+m_{\pi}\sqrt{s})][x-(m_{\pi}^2-m_{\pi}\sqrt{s})]}{[x-\frac{1}{2}
  (s+3m_{\pi}^2)]}.
\end{align}
There are two branches which we call $S$ and $H$, respectively. Each of these has $S^+$ and $S^-$ ($H^+$ and $H^-$) depending on whether $y>0$ or $y<0$. We have the mirror property
\begin{equation}
 \sigma^+(\sigma^*)= \sigma^+(\sigma)^*, \quad \sigma^-(\sigma^*)= \sigma^-(\sigma)^*. 
\end{equation}
We find that the various domains $u_{\pm}$, $v_{\pm}$, and $w_{\pm}$ map into each other:
\begin{equation}
\begin{array}{r@{\quad}rcl@{\qquad\qquad}r@{\quad}rcl}
\sigma^+: & u_{\pm} & \to & v_{\mp}, & \sigma^-: & u_{\pm} & \to & w_{\pm},\\
          & v_{\pm} & \to & u_{\mp}, &           & v_{\pm} & \to & w_{\mp},\\
          & w_{\pm} & \to & u_{\pm}, &           & w_{\pm} & \to & v_{\mp}.
\end{array}
\end{equation}
We further find that certain arcs map into each other:
\begin{equation}
\begin{array}{r@{\quad}rcl@{\qquad\qquad}r@{\quad}rcl}
\sigma^+: & L_{\pm} & \to & H_{\pm}, & \sigma^-: & L_{\pm} & \to & H_{\mp},\\
          & R_{\pm} & \to & S_{\mp}, &           & R_{\pm} & \to & S_{\pm},\\
          & H_{\pm} & \to & H_{\mp}, &           & H_{\pm} & \to & L,\\
          & S_{\pm} & \to & R,       &           & S_{\pm} & \to & S_{\mp}. 
\end{array}
\end{equation}
As the integration contour is rotated into the lower half-plane, the domain of analyticity of the unphysical sheet in the lower half-plane is exposed. In Fig.~\ref{fig:reference.plane}, the domain of analyticity in the lower half-plane is shown for a contour rotated by an angle $\phi$. Outside this domain, the contour $\sigma^-\to\sigma^+$ intersects the dispersive
integration contour $\sigma''$.
\begin{widetext}
Using relations in Eq.~\eqref{eq:comp-rela}, the $3\to 3$ scattering amplitude in Eq.~\eqref{eq:disprep-ful} can be rewritten as
\begin{align}
  T(\sigma'_+,s_+,\sigma_+) =& \int_{4m_{\pi}^2}^{+\infty}\frac{d\sigma''}{\sigma''-\sigma'-i\epsilon}\,\frac{6 s\, t^\dagger(\sigma''_+)t(\sigma_+)}{\lambda_{s}^{1/2}(\sigma'')\lambda_{s}^{1/2}(\sigma)} \,\theta^+(\phi(\sigma'',s,\sigma)) + \frac{1}{2\pi}\int_{4m_{\pi}^2}^{+\infty}\frac{d\sigma''} {\sigma''-\sigma'-i\epsilon}t(\sigma''_+)\rho(\sigma'') T(\sigma''_-,s_+,\sigma_+)\nonumber \\
 & +\frac{1}{\pi}\int_{4m_{\pi}^2}^{+\infty}\frac{d\sigma''} {\sigma''-\sigma'-i\epsilon}\frac{t(\sigma''_+)}{\lambda_{s}^{1/2}(\sigma'')} \frac{1}{8\pi} \int_{\sigma^-(\sigma'',s)}^{\sigma^+(\sigma'',s)} d\sigma_3 \, T(\sigma_{3}^*,s_+,\sigma_+). 
 \label{eq:T3to3}
\end{align}
\end{widetext}

Along the physical axis, $t(\sigma_+)$ and the analytic function $t^{\mathrm{II}}(\sigma_-)$ coincide with $t^{*}(\sigma_-)$ due to the Schwarz reflection principle and the continuity of the Riemann-sheet structure, since
\begin{align}
  t^{*}(\sigma_-)=t^{\mathrm{II}}(\sigma_-)=t(\sigma_+).
\end{align}
As the three-body amplitude is analytically continued into the lower half plane, the two-body subsystem lies on the
unphysical Riemann sheet. A pole on the unphysical Riemann sheet of the two-body subsystem requires a proper prescription for the dispersive integration contour. As the dispersive integration contour is rotated below the two-body pole position, as shown in the right panel of Fig.~\ref{fig:reference.plane}, the residue of the pole in the two-body interaction will be picked up. To ensure the amplitude is analytically continued, two terms need to be subtracted 
\begin{align}
  &-i\frac{1} {\sigma_\text{pole}-\sigma'}r_p\rho(\sigma_\text{pole}) T(\sigma_\text{pole},s_+,\sigma_+)\nonumber \\
   & -2i\frac{1} {\sigma_\text{pole}-\sigma'-i\epsilon}\frac{r_p}{\lambda_{s}^{1/2}(\sigma_\text{pole})} \nonumber \\
   &\times\frac{1}{8\pi} \int_{\sigma^-(\sigma_\text{pole},s)}^{\sigma^+(\sigma_\text{pole},s)} d\sigma_3 \, T(\sigma_{3}^*,s_+,\sigma_+),
\end{align}
where $r_p=\operatorname{Res}_{z=\sigma_\text{pole}}t^{\mathrm{II}}(z)$ is the residue of the simple two-body pole.
Associated with the two-body pole, there exists a complex two-body cut in the three-body scattering amplitude. The singularities can be obtained by specifying the dispersive integration contour and setting 
\begin{align}
\sigma_\mathrm{II}^{\pm}(\sigma'',s)=\sigma_\mathrm{pole},  
\end{align}
where $\sigma_\mathrm{II}^{\pm}(\sigma'',s)$ are the parts of the integration contour $\sigma^{\pm}(\sigma'',s)$ that cross the unitarity cut. The complex two-body cut is obtained by taking the upper boundary of the singularities.
For the $3\to 3$ scattering amplitude with a pair-wise interaction in Eq.~\eqref{eq:disrep-pair}, the dispersive integration contour will not intersect with the pole from the two-body interaction, and only $\sigma_\mathrm{II}^{\pm}(\sigma'',s)$ will intersect with the pole from the two-body interaction.

\section{Summary}
\label{sec:summary}

The present work illustrates the basic ideas for constructing the $3\to 3$ scattering amplitude from dispersion relations by using the subenergy as the dispersive integral variable. It is shown that three-body unitarity is automatically generated by two-body unitarity, analyticity and crossing symmetry. In this framework, requirements of two-body unitarity,
three-body unitarity, analyticity, and crossing are simultaneously satisfied. 
We have specified an analytical continuation of the $3\to 3$ scattering amplitude into unphysical energy sheets adjacent to the real axis
of the physical sheet.

For a pair-wise parametrization of the two-body subsystems, using Cauchy's theorem we have shown that the dispersive representation can be rewritten in a form closely related to the isobar--spectator formulations in Refs.~\cite{Freedman:1966xx,Aaron:1968aoz,Mai:2017vot,Jackura:2018xnx}. The resulting equation has the driving-term-plus-rescattering structure familiar from Bethe--Salpeter and Faddeev-type integral equations.

\begin{acknowledgments}
  This work is supported in part by the National Key R\&D Program of China under Grant No. 2023YFA1606703; by the National Natural Science Foundation of China (NSFC) under Grants No. 12125507, No. 12361141819, No. 12447101, No. 12405106 and No. 12247139; and by the Chinese Academy of Sciences (CAS) under Grant No.~YSBR-101.
\end{acknowledgments}
\bibliography{ref.bib}
\end{document}